\documentclass[10pt,journal,compsoc]{IEEEtran}

\usepackage{hyperref}
\usepackage{arydshln}
\usepackage{amsmath}
\usepackage{float}
\usepackage{multirow}
\usepackage{graphicx}
\usepackage{color}

\ifCLASSOPTIONcompsoc
  \usepackage[nocompress]{cite}
\else
  \usepackage{cite}
\fi

\ifCLASSINFOpdf
\else
\fi

\begin{document}
\title{Semi-Supervised Learning and Data Augmentation in Wearable-based Momentary Stress Detection in the Wild}

\author{Han~Yu,~\IEEEmembership{Member,~IEEE,}
        Akane~Sano,~\IEEEmembership{Member,~IEEE}
\IEEEcompsocitemizethanks{\IEEEcompsocthanksitem H. Yu and A. Sano are with the Department
of Electrical and Computer Engineering, Rice University, Houston,
TX, 77005.\protect\\
E-mail: see https://compwell.rice.edu
}
\thanks{Manuscript received April 19, 2005; revised August 26, 2015.}}

\IEEEtitleabstractindextext{%
\begin{abstract}
Physiological and behavioral data collected from wearable or mobile sensors have been used to estimate self-reported stress levels. Since the stress annotation usually relies on self-reports during the study, a limited amount of labeled data can be an obstacle in developing accurate and generalized stress predicting models. On the other hand, the sensors can continuously capture signals without annotations. This work investigates leveraging unlabeled wearable sensor data for stress detection in the wild. We first applied data augmentation techniques on the physiological and behavioral data to improve the robustness of supervised stress detection models. Using an auto-encoder with actively selected unlabeled sequences, we pre-trained the supervised model structure to leverage the information learned from unlabeled samples. Then, we developed a semi-supervised learning framework to leverage the unlabeled data sequences. We combined data augmentation techniques with consistency regularization, which enforces the consistency of prediction output based on augmented and original unlabeled data. We validated these methods using three wearable/mobile sensor datasets collected in the wild. Our results showed that combining the proposed methods improved stress classification performance by 7.7\% to 13.8\% on the evaluated datasets, compared to the baseline supervised learning models.

\end{abstract}

}

\maketitle

\IEEEdisplaynontitleabstractindextext

%
\IEEEpeerreviewmaketitle

\IEEEraisesectionheading{\section{Introduction}\label{sec:introduction}}
\IEEEPARstart{S}{tress} is common and complex. It can benefit people under certain circumstances and increase resilience. Exposure to moderate levels of stress can be beneficial as it can prepare an organism to deal with challenges \cite{dhabhar2014effects}. On the other hand, stress has also been associated with an increased risk for many somatic and mental illnesses \cite{aschbacher2013good}, increasing risks for cardiovascular health issues \cite{kario2003disasters} and suppressing the human immune system \cite{khansari1990effects}. Effectively detecting moments of stress in real life may help an individual regulate their stress behaviorally to promote resilience and wellbeing.

Widespread portable devices bring potentials in measuring human emotion, including stress levels, using passively sensed data. For example, wearable sensors and smartphones have enabled to monitor physiological and behavioral data such as body acceleration, skin conductance, skin temperature, heart rate, and phone usage data in real-time. Prior studies have developed machine learning models to measure momentary self-reported stress levels using physiological and behavioral sensor and survey features \cite{sanchez2017towards, gjoreski2017monitoring, shi-stress, ciman2016individuals}. For example, a 3-class momentary stress classifier (low/med/high) was developed with the highest accuracy of 94\% using behavioral and physiological wearable data and online survey data and Adaboost \cite{sanchez2017towards}. A support vector machine (SVM) based binary momentary stress detector was designed with 70\% recall and 95 \% precision rates \cite{gjoreski2017monitoring}. Moreover, since physiological and behavioral sensor data are naturally time sequences, researchers applied time-series models to boost the model performances. For example, long short-term memory (LSTM) network has been used in estimating stress \cite{umematsu2019improving}. Furthermore, an attention based LSTM structure was applied to multimodal data, including physiological signals, to infer human emotion \cite{yang2021behavioral}. In this work, we focus on estimating self-reported momentary stress with time-series models and wearable/mobile sensor data. 

Although these prior studies provided promising results in stress estimation, we could further improve the performances by addressing challenges in data. In the studies introduced above, the number of self-reported labels is usually limited. On the other hand, wearable devices can collect millions of data samples throughout the study period. The aforementioned studies focused on using data aligned with stress labels, which resulted in information loss when discarding unlabeled data. In alleviating this issue, various semi-supervised algorithms, which used both labeled data and unlabeled data during machine learning training, have been applied to improve stress estimation model performance. For instance, Maxhuni \textit{et al.} designed a semi-supervised ensemble learning method in binary momentary stress detection and achieved an f1 score of 0.70 using mobile phone data \cite{maxhuni2016stress}. Semi-supervised sequential learning has been developed rapidly and performed well in other areas such as natural language processing (NLP) \cite{dai2015semi} and human cardiovascular risk detection \cite{ballinger2018deepheart}. Nevertheless, improvement in model performance cannot be guaranteed by directly applying methods in the other fields. For example, the differences in the distribution of labeled and unlabeled data might hinder learning representations using the method in \cite{dai2015semi}. The methods mentioned above might not be able to apply to wearable or mobile sensor data collected in the wild as they are noisy and using all of the unlabeled data might not be effective for learning robust models.

Thus, in this work, we proposed novel hinges for effectively learning stress detection models by actively sampling unlabeled sensor data and augmenting unlabeled and labeled sensor data. The new framework showed improvements in human momentary stress estimation using physiological and behavioral data collected in the wild. 

Our contributions are summarized as:
\begin{itemize}
    \item We proposed a semi-supervised active sampling method for selecting unlabeled data adaptively based on the distributions of labeled data. 
    \item We developed data augmentation and consistency regularization approaches for both labeled and unlabeled data for robust model training.
    \item We evaluated the proposed methods using three datasets, including multimodal sensor data and momentary stress labels collected in the wild. We observed the improvements in model performances using the proposed methods. 
\end{itemize}

\section{Related Work}
Traditionally, stress was measured using surveys. For example, the perceived stress scale (PSS) was developed for measuring the perceived stress over the past month \cite{PSS}. The Holmes and Rahe Stress Scale is another survey instrument that adds up the self-reported events in the prior year that could lead to stress, estimates the total amount of stress in the past year, and determines whether people are at risk of becoming sick \cite{Social1967}. However, a drawback of these traditional surveys is that they might be cumbersome for individuals to complete, especially when the studies last for a long time. The resulting fatigue may make survey answers less reliable.

With the development of mobile phones and wearable devices, accessing users' physiological and behavioral data in daily life settings has become a boost in monitoring human mental status. 
Machine learning has enabled us to develop models to learn patterns from data samples and have already brought benefits in ubiquitous computing applications. Multi-modal data from wearable sensors, mobile phones, and other smart devices have been widely used with machine learning in estimating human emotion as well as momentary stress levels\cite{sanchez2017towards, gjoreski2017monitoring, shi-stress, yang2021behavioral, ciman2016individuals, hinkle2019physiological}.
Yang \textit{et al.} proposed an attention-based LSTM system to fit data from smartphones and wristbands and predicted the participants' positive or negative emotional states with an accuracy of 89\%\cite{yang2021behavioral}.
Hinkle \textit{et al.} leveraged multi-modal physiological signals - such as electrocardiogram (ECG) and Electroencephalogram (EEG) - to detect human emotions by classifying binary classes of arousal and valence \cite{hinkle2019physiological}. They achieved an accuracy of 89\% with an SVM model.
Sanchez \textit{et al.} used Fitbit wrist-worn devices and online surveys to record the physiological and behavioral data of 41 employees from 3 companies \cite{sanchez2017towards}. Study participants filled out surveys twice a day to monitor momentary stress levels, and those labels were divided into three classes (low/med/high) as target labels. The authors then balanced the training set with an oversampling approach and achieved 94\% estimating accuracy with machine learning models such as SVM, logistic regression, etc. Shi \textit{et al.} collected 22 subjects' ECG, galvanic skin response (GSR), respiration (RIP), and skin temperature (ST) data using wearable sensors \cite{shi-stress}. Each subject in the study was exposed to a protocol, including four stressors and six rest periods, and stress labels were collected before and after each stressor/rest period through interviews. The authors proposed a personalized an SVM algorithm to classify binary stress labels (low/high), which provided 0.68 precision with 0.80 recall.
In these studies, while physiological data were collected continuously using wearable sensors, human stress labels were collected using questionnaires. 
To match the sensor data with sparse labels, the collected physiological data were downsampled to align momentary features with stress labels, which \cite{sanchez2017towards, shi-stress}, caused a loss of information. 

In order to overcome the difficulty in learning models with a small number of labels, semi-supervised learning methods have been applied in mobile and wearable device studies \cite{ballinger2018deepheart, maxhuni2016stress, song2017development}, where models learned representations from massive unlabeled data and used a small amount of labeled data to train the supervised prediction models. Ballinger \textit{et al.} applied a semi-supervised auto-encoder pretraining method \cite{dai2015semi} and used physiological data collected from wrist-wearable devices to detect symptoms of cardiovascular disease. Their semi-supervised model improved AUC scores for 3 out of 4 types of symptoms, and the highest improvement rate was 10.5\% in detecting high cholesterol. In momentary stress detection, Maxhuni \textit{et al.} used a self-training tree model in combining unlabeled wearable sensor-based physiological data with labeled data \cite{maxhuni2016stress}. Compared to the supervised learning method, their binary stress detection performance in the f1 score boosted from 66.0\% to 70.0\%.
In our work, we used semi-supervised sequential learning to extract representations from data more effectively since our target physiological data are sequential.

\section{Methods}
\label{methods}
\begin{figure*}
	\centering
	\includegraphics[scale=0.34]{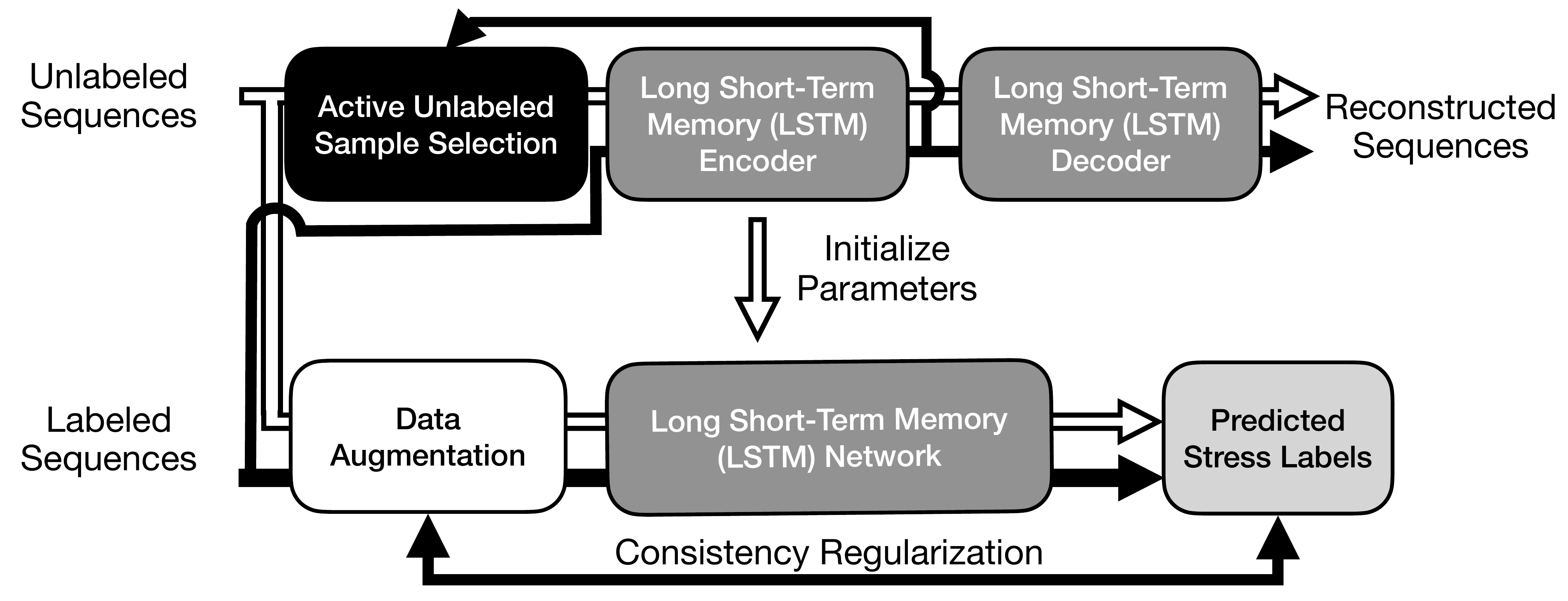}
	\caption{The overall structure of the designed semi-supervised sequence learning framework for stress estimation. The framework contains the components of data augmentation, LSTM auto-encoder pre-training with active unlabeled sample selection, and consistence regularization.}
	\label{overall_framework}
\end{figure*}

This section introduces our proposed semi-supervised learning method for leveraging both labeled and unlabeled sequences in stress estimation. Figure \ref{overall_framework} shows the overall framework of the designed approach including data augmentation on physiological and behavioral sequences, unsupervised LSTM auto-encoder pre-training with active sampling, and consistency regularization. Finally, we also introduce the approach we used for interpreting the stress detection model.

\subsection{Data Augmentation}
\label{DA}
Training a deep neural network model requires a large amount of data for learning robust model parameters; however, obtaining labels is expensive and we usually have access to only limited labeled datasets. Data augmentation (DA) synthesizes a comprehensive set of possible data points and could alleviate the limited data problems. Further, DA approaches can generate transformed samples even for unlabeled data, which is the foundation of the consistence regularization method we introduces in section \ref{consistency_reg}. 

\begin{figure}
	\centering
	\includegraphics[scale=0.34]{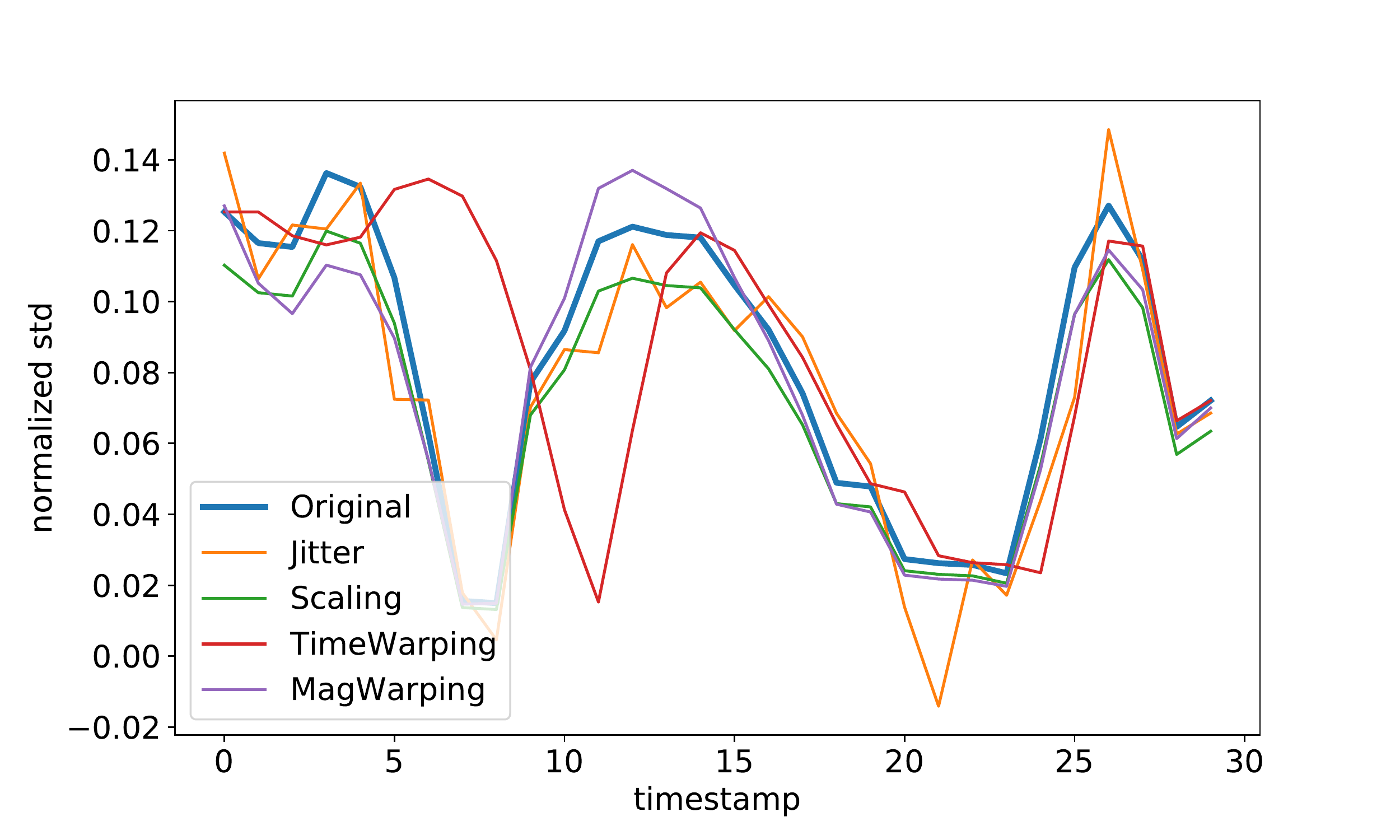}
	\caption{Examples of data augmentation on a sequence of ACC std signal}
	\label{example_da}
\end{figure}

We extended 4 DA methods from \cite{um2017data} to physiological and behavioral sequences. These 4 methods are jittering, scaling, time warping, and magnitude warping. Jittering (J) added tiny Gaussian noise to the original signals. For scaling (S), the original signals are scaled by generated Gaussian random numbers ($\mathcal{N}\sim(1, 0.05)$). Time warping (TW) is a way to perturb the temporal characteristics of the data. The temporal locations of the samples are changed by smoothly distorting the time intervals between samples.
Magnitude warping (MW) changes the magnitude of each sample by convoluting the data window with a smooth curve varying around one with a standard deviation of 0.05 ($\mathcal{N}\sim(1, 0.05)$). The essence of these methods is adding a small amount of noise to time-series data so that the trained model will be robust. Figure \ref{example_da} shows an example of different DA methods on a 30-minute sequence of accelerometer standard deviation data. The blue line is the original signal, and other lines  represent the data generated using 4 different DA methods. We applied the combination of all these methods in our experiments.

\subsection{Baseline Model: Long Short-Time Memory (LSTM) network}
\label{baseline_LSTM}
Long short-term memory (LSTM) networks \cite{LSTM}, as an extended type of recurrent neural networks, and have been used in time-series applications \cite{selvin2017stock, cho2014learning}. In some previous studies, LSTM models provided promising results in stress regression with time-series sensor data\cite{Previous-reg2, MTL-LSTM-KDD}.

Considering that people's current stress status might be affected by previous short-term physiology or behavior changes, we applied each participant's previous time steps of the data to a multi-layer LSTM for sequential learning. Moreover, we found the distributions of the training data might vary among participants. For example, different participants might have different average heart rate, which introduced the internal covariate shifts among samples. Thus, we also applied a batch normalization (BN) layer\cite{BN} after LSTM so that the high-level temporal features extracted by LSTM would be scaled and shifted into the same distribution.

\subsection{Semi-Supervised Active Learning}
\subsubsection{Semi-Supervised Sequence Learning}
\label{semi_seq_learning}
Semi-supervised sequence learning, which uses a sequence-to-sequence auto-encoder in learning representations from unlabeled data, has been shown to improve the model performance when there was a large amount of unlabeled training data\cite{dai2015semi}. Ballinger \textit{et al.} achieved improvements using this method compared to their baseline model when predicting cardiovascular risks with physiological data\cite{ballinger2018deepheart}. Inspired by the previous work, we experimented with this approach by ﬁrst constructing a sequence-to-sequence LSTM auto-encoder (LSTM-AE). The structure of the encoder and decoder of the LSTM auto-encoder followed the 
exact structure of the baseline LSTM model in section \ref{baseline_LSTM}. The input $X$ of LSTM-AE were the time-series physiological feature sequences, then the output of the decoder returned the reconstructed sequences $\hat{X}$. The loss function was the mean square loss between the original sequence $X$ and the reconstructed output $\hat{X}$:
\begin{equation}
    L_{AE} = \|X - \hat{X}\|^2
\end{equation}

\begin{figure}
	\centering
	\includegraphics[scale=0.5]{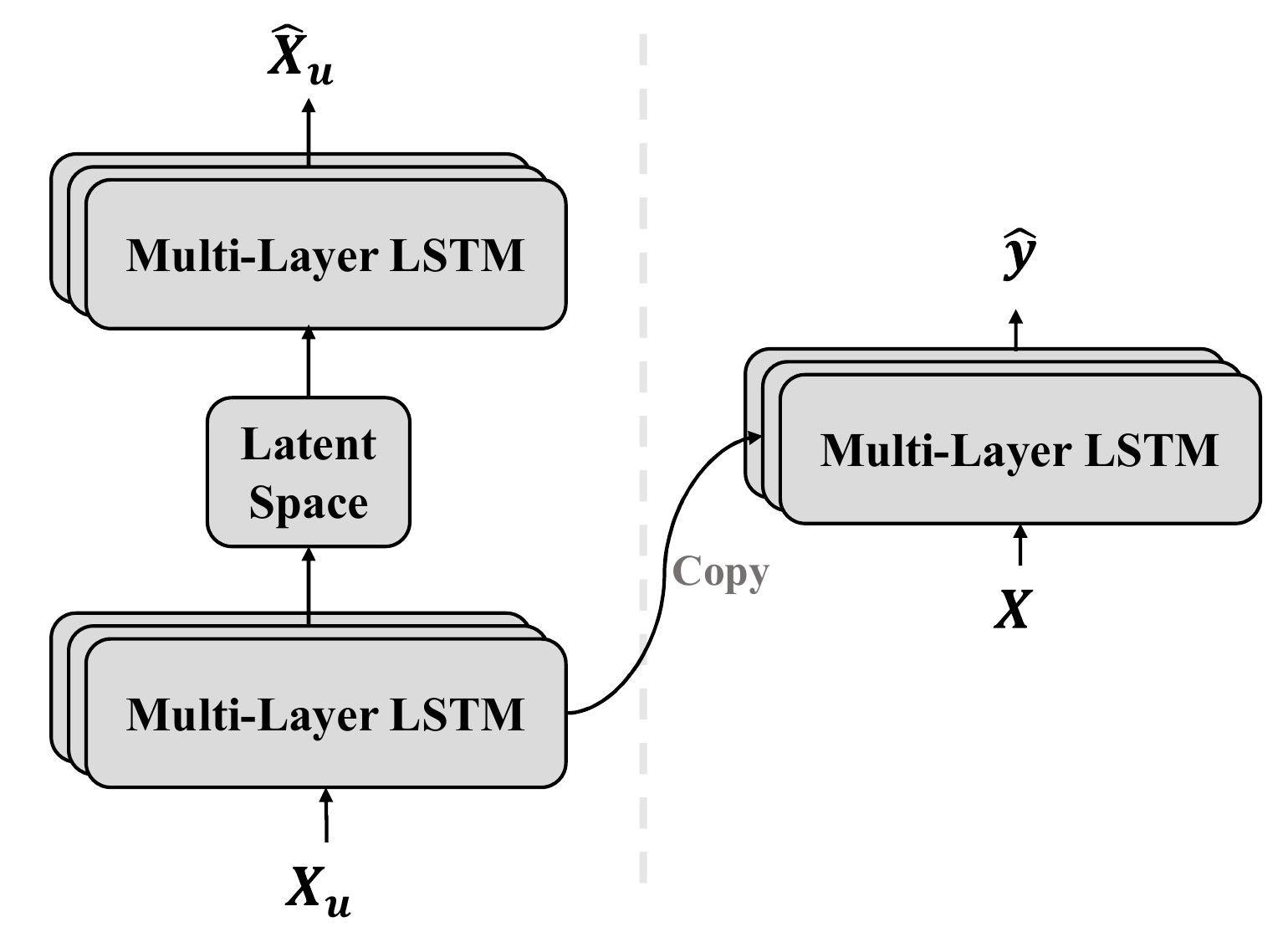}
	\caption{The structure of long short-term memory auto-encoder pre-training method. The left side LSTM layers are optimized by reducing the loss between the unlabeled input sequence $X_u$ and reconstructed $\hat{X_u}$. Next, the trained LSTM layers' parameters are used to initialize weights in our baseline LSTM layers. Then, our baseline LSTM model is trained through the labeled data sequence $X$, and outputs estimated stress labels $y$.}
	\label{LSTMAEstructure}
\end{figure}

After the LSTM-AE was trained, we used the parameters of the LSTM-AE encoder layers as the initial parameters for the corresponding layers in the supervised architecture. The mechanism of this method is shown in Figure \ref{LSTMAEstructure}. Moreover, in order to enhance the robustness of the model, we added the Gaussian noise to the input of encoder when training an LSTM-AE model.

\subsubsection{Active Unlabeled Sample Selection}
Wearable physiological signals collected in the wild contain noise. In addition, human physiological signals can vary widely in different states. These noise and uncertainty can cause significant distribution differences between labeled and unlabeled data samples as $r$.  The distribution of the unlabeled data is usually more comprehensive, and a significant fraction of them might even distribute differently from the labeled data.

\begin{figure}
	\centering
	\includegraphics[scale=0.6]{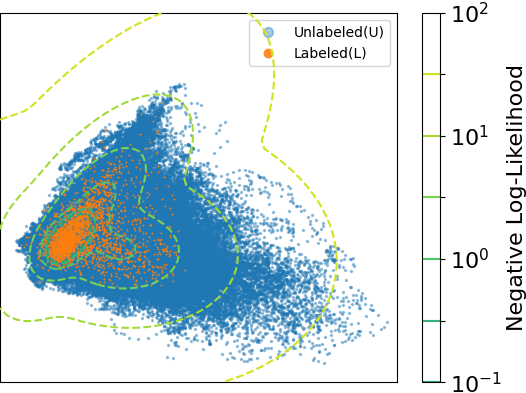}
	\centering
	\caption{Latent space PCA-based low dimension mapping visualization. The representations of labeled samples are highlighted in orange color. Example visualization in the SMILE dataset with 3 gaussian mixture components.}
	\label{latent_space_viz}
\end{figure}

To reduce the influence of noise and unlabeled samples with different distributions on the LSTM-AE pre-trained parameters of the model, we propose an active unlabeled sample selection method. We first trained an LSTM-AE with labeled data only, then clustered all labeled samples in latent space low-dimension representation using a Gaussian mixture model (GMM). After analyzing the elbow points of both the Akaike and Bayesian information criterion, we fixed the number of Gaussian components as K. Then, we used the trained encoder to infer the latent representations of all the unlabeled samples as $h(x_u)$. The negative log-likelihood of each unlabeled samples, which is the probability of the observed data under the trained GMM model, can be calculated via:
\begin{equation}
    \ell(\mathbf{x_u})= - \log\left( \sum_{m=1}^{K} \alpha_m \phi(\mathbf{h(x_u)}|\mu_m,\Sigma_m)\right)
\end{equation}
where $\alpha$ represents the weight mixture component, $\mu$ and $\Sigma$ are the learned mean value and co-variance of the corresponding Gaussian component.

Then, we selected the unlabeled samples that most possibly obey the similar distributions of labeled samples based on the calculated negative log-likelihood values (NLL). The smaller the NLL, the more similar the sample distributed as labeled data. Figure \ref{latent_space_viz} shows the reduced-dimensional visualization of LSTM-AE latent space representations, and the contour lines indicate the negative log-likelihood levels across the whole dataset. Then, we trained LSTM-AE with only the selected unlabeled samples. Under this scenario, the pre-trained model can focus on the information learned from samples with a similar distribution as the labeled data. 

\subsection{Consistency Regularization}
\label{consistency_reg}
Consistency training methods regularize model predictions to be invariant to slight noise applied to input \cite{miyato2018virtual, xie2019unsupervised}. The theoretical foundation of this method is that a robust machine learning model should be able to tolerate any slight noise in an input example. For example, when inputting a data sequence and its augmented sequence into a robust model, the outputs of those two input examples should be the same. Since there were plenty of unlabeled sequences in our dataset, ideally, the concept of consistency regularization could bring robustness to our model. 

In our study, inspired by \cite{xie2019unsupervised}, we conducted consistency training combined with the augmented data discussed in section \ref{DA}. We generated $M$ new augmented sequences using each labeled/unlabeled time window. 
If there are any differences between the LSTM model outputs based on the original labeled data and their augmented input data, the consistency losses will be regularized to the supervised loss function. For example, since our task was to estimate stress status in binary classification, the supervised loss would be a cross-entropy loss. We applied the Kullback-Leibler divergence loss as our designed consistency loss. To present the method in formula, the final loss function with the consistency regularization method is:

\begin{equation}
\begin{split}
    L = &L_{CE}(X_{l}, y) + \frac{\alpha}{M}\sum_{m=1}^{M} L_{KL}(p(y_{l} | X_{l}), p(y_{l} | \bar{X_{l}^{m}})) \\
    & + \frac{\lambda}{M}\sum_{m=1}^{M} L_{KL}(p(y_{u} | X_{u}), p(y_{u} | \bar{X_{u}}^{m}))
\end{split}
\end{equation}

where $X_l$ and $y$ represent the labeled data and ground truth labels, $X_u$ is the unlabeled sequence, $\bar{X_{l}}$ is the augmented labeled sequence, and $\bar{X_{u}}$ is the augmented unlabeled sequence. $y_l$ and $y_u$ represent the output from the model using the given input data sequence $X_l$ and $X_u$, respectively. The probability $p(y|x)$ indicates the likelihood of getting model results with given data $x$. In our case of classification, $p(y|x)$ is the sigmoid outputs for binary classification. $\alpha$ and $\lambda$ control the weights of the consistency regularization, which are both set to 1 in this study. 


\subsection{Saliency Map}
\label{method:saliency_map}
Our proposed semi-supervised learning framework uses deep neural networks such as the auto-encoder and LSTM. Understanding the relationship between input and output in these deep learning components is complicated since massive non-linear operations are involved. On the other hand, we believe that explainability is important in our stress estimation model. By interpreting the model parameters, we would learn the factors that influencing participants' stress levels and trigger proper interventions to help relieve stress. Therefore, we used saliency maps \cite{simonyan2013deep} to interpret the importance of our input signals in stress detection. The saliency map uses the gradient of a neural network model and finds the input signal's salient region. As the output of the saliency map, higher output values indicate that the corresponding input has considerable importance to the model. We calculated the average saliency of all input samples to understand the overall importance of features and time.

\section{Datasets}
In this section, we describe three datasets we used to evaluate our methods. \label{stress_datasets}
\subsection{Dataset I: SMILE}
Wearable sensor and self-report data were collected from 45 healthy participants (39 females and 6 males), in total for 390 days. The average age of participants was 24.5 years old, with a standard deviation of 3.0 years. Participants contributed to an average of 8.7 days of data, with a minimum of 5 days and a maximum of 9 days.

Two types of wearable sensors were used for data collection \cite{smets2018towards}. One was a wrist-worn device (Chillband, IMEC, Belgium) designed for the measurement of skin conductance (SC), ST, and acceleration data (ACC). The SC was sampled at 256 Hz, ST at 1 Hz, and ACC at 32 Hz. Participants wore the sensor for the entire testing period, but could take it off during the night and while taking a shower or during vigorous activities. The second sensor was a chest patch (Health Patch, IMEC, Belgium) to measure ECG and ACC. It contains a sensor node designed to monitor ECG at 256 Hz and ACC at 32 Hz continuously throughout the study period. Participants could remove the patch while showering or before doing intense exercises.

\subsubsection{Stress Labels}
In addition to the physiological data collected by sensors, participants received notifications on their mobile phones to report their momentary stress levels 10 times per day, spaced out roughly 90 minutes apart for eight consecutive days. In total, 2494 stress labels were collected across all participants (80\% compliance). The stress scale ranged from 1 ("not at all") to 7 ("Extreme"). 
In 45\% of the cases, participants reported that they were not under stress, while in only 2\% of the cases did they report that they were under extreme stress. In this work, we binarized the stress levels by categorizing stress level 1 as a class of "non-stressed" (45\%) and level 2-7 as the "stressed" class (55\%).


\subsection{Dataset II: TILES}
Tracking Individual Performance with Sensors (TILES) is a multi-modal data set for the analysis of stress, task performance, behavior, and other factors pertaining to professionals engaged in a high-stress workplace environment \cite{mundnich2020tiles}. The dataset was collected from 212 participants for 10 weeks. In this work, we leveraged the ECG signals, which were not collected in a strictly continuous manner. At five-minute intervals, the sensor collects ECG signals for fifteen seconds at a sampling rate of 250 Hz for the participants. We extracted features using the ECG signals and estimated the self-reported stress levels.

Gaballah \textit{et al.} leveraged TILES audio and physiological data with a bidirectional LSTM network and inferred stress labels in a binary classification task\cite{gaballah2021context}. They achieved a f1-score of 0.64. With extracted features from ECG signals, Pimentel \textit{et al.} proposed SVM based binary stress detection models with an f1-score of 0.68 \cite{pimentel2021human}.

\subsubsection{Stress Labels}
Participants annotated stress levels through multiple 5-point scale questions. Following the stress label processing procedures in \cite{gaballah2021context}, 
We calculated the z-scores of stress levels for each individual, considering the subjective variability  and then divided them into two classes, class 0 (non-stressed, z-score below the average) and class 1 (stressed, z-score above the average). Overall, 600 stressed labels and 629 non-stressed labels were processed.

\subsection{Dataset III: CrossCheck}
We also used the dataset from the CrossCheck project \cite{wang2016crosscheck}, where smartphone data were collected in a clinical trial from 75 patients with schizophrenia for up to a year. The collected phone data include acceleration, light levels, sound levels, GPS, and call/SMS meta data. In addition, the participants filled out the momentary ecological assessment (EMA) up to 3 times a week to assess their symptoms.

Prior work estimated schizophrenia symptoms including depression, harm, stress, etc using machine learning regression models and behavioral features extracted from the phone data. \cite{wang2016crosscheck}. The best mean absolute error performance from the non-personalized model was 1.5 out of 0-3 scale. Similar symptom estimation performance was also reported in \cite{tseng2020using, wang2017predicting}.

\subsubsection{Stress Labels}
Stress labels were collected via EMA. Participants reported their stress levels using a 4-point scale, where "0" means no stress at all, whereas "3" means extremely stressed. The total number of stress labels was 5914, where 49\% was "0" stress label class, 28\% was "1", 16\% was "2", and 7\% was "3". We evaluated the proposed methods in a binary classification task on "0" (non-stressed) versus "1, 2, 3" (stressed). 

\subsection{Features Extraction}
We used features extracted from the period prior to the stress label to develop stress detection models and learn the temporal representations.
In the three datasets mentioned in section \ref{stress_datasets}, ECG (SMILE, TILES), SC (SMILE), ST (SMILE), ACC (SMILE) and smartphone logs (CrossCheck) were used in estimating stress level. All these signals contribute to infer human stress. For example, ECG reflects sympathetic and parasympathetic activity, which has been proven related to stress \cite{kim2018stress,sharma2012objective}. Stress has also been proven related to conductance in human eccrine sweat glands, which can be captured by sensors as SC \cite{harrison2006skin}.
ST changes could reveal the intensity of stress \cite{herborn2015skin}. ACC is associated with human physical activity, which has been shown influenced by stress level \cite{stults2014effects}. Also, previous studies have shown that smartphone usage data contributed to stress prediction \cite{sano2018identifying, umematsu2019improving}. 

We calculated statistical features from ST and ACC data, such as mean and standard deviation values across periods. For other sensor measurement,  we introduce the features extracted from ECG, SC, and smartphone data in this section.

\subsubsection{ECG Features}
To extract features, based on the data sampling pattern of sensors in different studies, we used 15 seconds of high-resolution ECG data every five minutes for the TILES dataset and continuous ECG data for the SMILE dataset. We extracted both time and frequency domain ECG features using a Python library \cite{hrvanalysis}. The engineered time-domain features covered subjects' heart rate, heart variability, etc. The frequency-domain features contained the power spectrum information of heart activities in various frequency bands. These features has been proven associated with human stress levels \cite{kim2018stress, sharma2012objective}. The detailed list of ECG features is available in Appendix \ref{appen:ECG_fea_list}.


\subsubsection{SC Features}
We computed SC magnitude, number of SC responses, the response duration, etc, following previous stress studies \cite{kim2018stress}. See the detailed list of SC features in Appendix \ref{appen:SC_fea_list}.

\subsubsection{Smartphone Features Extraction}
We processed the CrossCheck data and extracted features using data collected by smartphones. These features include acceleration intensity, phone application usage, call/sms counts and duration, and the location related features from GPS. See the detailed list of smartphone features in Appendix \ref{appen:smartphone_fea_list}.

\section{Experiments}
\subsection{Model Parameters}
We conducted experiments in a participant-independent setting, where we conducted a 5-fold cross-validation for each dataset by splitting data from 80\% of the participants as training sets and the rest as validation sets. For example, in the SMILE dataset with 45 participants, we selected data from 10 participants as a validation set for each cross-validation. This way, the models did not see data from the participants in the training set during each validation. 

For hyperparameters of the baseline LSTM model, we used grid searching through cross-validation. We adopted the following structure and parameters: 
\begin{itemize}
    \item \textbf{SMILE}: 3 layers of LSTM with 64 recurrent units were connected, incorporating 0.4 recurrent dropout and 0.3 dropout rates in each LSTM layer. After a BN layer, a fully-connected layer followed with 512 hidden units with a dropout rate of 0.5. We chose Adam as the optimizer with a learning rate of 0.0001.
    \item \textbf{TILES \& CrossCheck}: 3 layers of LSTM with 32 recurrent units were connected, incorporating 0.3 dropout rates in each LSTM layer. After a BN layer, a fully-connected layer followed with 256 hidden units with a dropout rate of 0.5. We chose Adam as the optimizer with a learning rate of 0.00005.
\end{itemize}

\subsection{Comparison of different feature extraction windows and input sequence lengths}
To compare the influence of feature extraction window and input sequence lengths on the model performances, We compared various combinations of feature resolution and sequence lengths. 
 We compared the resolution of feature extraction, every [1, 5, 10, 20] minute. In the TILES dataset, the ECG data was collected from participants every 5 minutes, thus we could not extract features with a resolution of every 1 minute. Similarly, the smartphone application usage data were logged every 20 minutes, thus we could not extract the APP usage features for resolution less than 20 minutes. For all three datasets, we compared the length of the sequences from 5 steps to 60 steps. 
\subsection{Data Segmentation}
After comparing different feature extraction window and sequence lengths as shown in the previous subsection, we used the length and resolution of the sequences that showed the best performance in the baseline supervised learning model for the rest of the experiments. 

\subsubsection{SMILE}
Each participant’s data were split into 30-step sequences, and each step represented features extracted from a data period of 5 minutes. Any time windows with missing data points were omitted. The size of each sequence was 30 $\times$ 37 (number of features). After time sequence segmentation, there were 2494 labeled sliding time windows (0.7\%) and 362519 unlabeled time windows (99.3\%), respectively. 

\subsubsection{TILES}
Each participant’s data were split into 30-step sequences, and each step represented features extracted from a data period of 5 minutes. The size of each sequence was 30 $\times$ 23 (number of features). The ratio of labeled sequences versus unlabeled sequences was 1229 (0.4\%) versus 349211 (99.6\%).

\subsubsection{CrossCheck}
Each participant’s data were split into 30-step sequences, and each step represented features extracted from a data period of 20 minutes. The size of each sequence was 30 $\times$ 18 (number of features). There were 2371 (0.4\%) labeled sequences and 441325 (99.6\%) unlabeled sequences in total.

\subsection{LSTM-AE Pre-Training with Active Sampling}
In the methods section \ref{methods}, we proposed an active sampling strategy to select unlabeled data for pre-training based on likelihood scores of following the distribution of labeled samples. This experiment controlled different volumes of unlabeled samples according to the latent representation distribution to pre-train the model. Therefore, we compared the model performances based on NLL scores we threshold the samples selected. To verify the effectiveness of active unlabeled sample selection, we also evaluated the pre-trained models with randomly sampled data and compared the results with our active sampling method.

\subsection{Semi-Supervised Learning Methods}
We introduced a variety of methods to improve model performance. To evaluate their respective contributions and the performance improvement, we conducted the experiments of evaluating the performances of the following models/method: (a) baseline LSTM model, (b) baseline LSTM with LSTM-AE pretraining, (c) baseline LSTM model with CR, and (d) baseline LSTM with both LSTM-AE pretraining and CR. To verify that the models successfully learned patterns, we also compared the performances of the baseline LSTM with the random baseline. Random baseline is the method that assigns labels to test instances according to the class probabilities in the training set \cite{bishop2006pattern}. For example, in the SMILE dataset, the probability of class 0 $p(y=0) = 0.45$, we assigned the instances in the test set as class 0 with the probability of 0.45. 

\subsection{Model Performance Analysis on Stress Sub-Levels}
To understand the binary stress detection performance differences with different originally reported stress levels, 
we analyzed the model performances by investigating the model accuracy on each of the stress sub-levels. Since our model conducts binary classifications, we counted the classification accuracy at the original stress sub-levels reported by the participants. For example, among 55\% of samples categorized as stressed in the SMILE dataset,  36\% of them was stress level 2 (lightly stressed) and 3\% of them was 7 (extremely stressed). We calculated the binary model performance for each stress sub-level separately. 

\subsection{Input Importance Visualization Using Saliency Map}
With the saliency map method we introduced in section \ref{method:saliency_map}, we visualized the overall model weights on the input samples. We showed the visualization results for all three datasets and found the most relevant elements in time series upon stress estimation.

\section{Results}
\subsection{Comparison of different input sequence lengths}
\label{compare_length}


Tables \ref{tab:supervised_tuning} shows the performances of the baseline LSTM model using the different combination of step size in minutes and the lengths of input sequences on the SMILE, TILES, and CrossCheck dataset, respectively.
In the SMILE dataset, we found the model performances with the sequence lengths of 30 and 60 with a step size of 5 minutes were significantly higher than the other combinations (ANOVA, p $<$ 0.01). Moreover, compared to the model performances with step resolution of 1 minute and 5 minutess, the one with 20-minute step resolution showed statistically lower performances for all sequence lengths (paired t-test, calibrated p $<$ 0.01).
In the TILES dataset, the models using 30/45/60 in sequence length with a 5-minute resolution performed significantly higher in f1-score compared to the other sequence lengths (ANOVA, p $<$ 0.01). Also, similarly to our observation in the SMILE dataset, the model with step resolution of 20 minutes performed lower in f1-score than the 5-minute resolution (paired t-test, p $<$ 0.01). 
In the CrossCheck dataset, we found that the model performed better using step sizes over 30 than shorter sequences when considering the features without app usage data. For example, the models built with a sequence length of 30 and resolution of 5 and 20 minutes showed statistically higher performances than using 5 or 15 steps correspondingly (ANOVA, Tukey, p $<$ 0.01). However, we did not observe significant differences in different resolutions when using sequence lengths over 30. Moreover, when using a sequence length of 30 with 20-minute resolution in the CrossCheck dataset, we found the model performance of using app usage data was significantly better than the one without using app data (paired t-test, calibrated p $<$ 0.01). Thus, in semi-supervised learning experiments on the CrossCheck dataset, we used sequences of 30 steps with a 20-minute resolution that covered app usage features.

\begin{table*}[]
\centering
\small
\caption{Baseline supervised LSTM performance for SMILE, TILES, and CrossCheck dataset with different combinations of step resolutions and sequence lengths. Step resolution is the the duration of the time period we used to extract the features for one step. No App means no app usage data were leveraged in the CrossCheck dataset. }
\label{tab:supervised_tuning}
\begin{tabular}{cc|ccccc|}
\cline{3-7}
                                                 &                                                                & \multicolumn{5}{c|}{Sequence Length}                                                                               \\ \cline{1-2}
\multicolumn{1}{c|}{Dataset}                     & \begin{tabular}[c]{@{}c@{}}Step Resolution \\ (minutes)\end{tabular} & 5               & 15              & 30                       & 45                       & 60                       \\ \hline
\multicolumn{1}{c|}{\multirow{4}{*}{SMILE}}       & 1                                                              & 0.52 $\pm$ 0.02 & 0.53 $\pm$ 0.01 & 0.56 $\pm$ 0.01          & 0.57 $\pm$ 0.02          & 0.57 $\pm$ 0.01          \\
\multicolumn{1}{c|}{}                            & 5                                                              & 0.53 $\pm$ 0.02 & 0.55 $\pm$ 0.02 & \textbf{0.58 $\pm$ 0.01} & 0.57 $\pm$ 0.01          & \textbf{0.58 $\pm$ 0.01} \\
\multicolumn{1}{c|}{}                            & 10                                                             & 0.52 $\pm$ 0.01 & 0.54 $\pm$ 0.01 & 0.57 $\pm$ 0.02          & 0.56 $\pm$ 0.01          & 0.56 $\pm$ 0.02          \\
\multicolumn{1}{c|}{}                            & 20                                                             & 0.50 $\pm$ 0.02 & 0.52 $\pm$ 0.03 & 0.53 $\pm$ 0.02          & 0.54 $\pm$ 0.01          & 0.54 $\pm$ 0.02          \\ \hline
\multicolumn{1}{c|}{\multirow{4}{*}{TILES}}      & 1                                                                                                                                            & -               & -               & -                                         & -                        & -                                         \\
\multicolumn{1}{c|}{}                            & 5                                                                                                                                            & 0.58 $\pm$ 0.02 & 0.63 $\pm$ 0.03 & \textbf{0.65 $\pm$ 0.02}                  & \textbf{0.65 $\pm$ 0.03} & \textbf{0.65 $\pm$ 0.02}                  \\
\multicolumn{1}{c|}{}                            & 10                                                                                                                                           & 0.59 $\pm$ 0.03 & 0.62 $\pm$ 0.02 & 0.64 $\pm$ 0.03                           & 0.63 $\pm$ 0.02          & 0.63 $\pm$ 0.02                           \\
\multicolumn{1}{c|}{}                            & 20                                                                                                                                           & 0.56 $\pm$ 0.02 & 0.58 $\pm$ 0.03 & 0.60 $\pm$ 0.03                           & 0.61 $\pm$ 0.02          & 0.61 $\pm$ 0.03                           \\ \hline
\multicolumn{1}{c|}{\multirow{5}{*}{CrossCheck}} & 1 (No App)                                                                                                                                   & 0.52 $\pm$ 0.03 & 0.55 $\pm$ 0.03 & 0.56 $\pm$ 0.04                           & 0.56 $\pm$ 0.05                           & 0.56 $\pm$ 0.04                           \\
\multicolumn{1}{c|}{}                            & 5 (No App)                                                                                                                                   & 0.53 $\pm$ 0.04 & 0.55 $\pm$ 0.03 & 0.57 $\pm$ 0.04                           & 0.56 $\pm$ 0.03                           & 0.57 $\pm$ 0.03                           \\
\multicolumn{1}{c|}{}                            & 10 (No App)                                                                                                                                  & 0.53 $\pm$ 0.03 & 0.55 $\pm$ 0.05 & 0.56 $\pm$ 0.04                           & 0.56 $\pm$ 0.03                           & 0.57 $\pm$ 0.03                           \\
\multicolumn{1}{c|}{}                            & 20 (No App)                                                                                                                                          & 0.53 $\pm$ 0.03 & 0.54 $\pm$ 0.04 & 0.57 $\pm$ 0.05                  & 0.57 $\pm$ 0.04                           & 0.56 $\pm$ 0.03                           \\ \cdashline{2-7}
\multicolumn{1}{c|}{}                            & 20 (With App)   & 0.55 $\pm$ 0.03 & 0.57 $\pm$ 0.03 & \textbf{0.59 $\pm$ 0.03}                  & 0.58 $\pm$ 0.02                           & 0.58 $\pm$ 0.03                           \\
\cline{2-7} \hline
\end{tabular}
\end{table*}

\subsection{LSTM-AE Pre-Training with Active Sampling}
\begin{figure*}
	\centering
	\includegraphics[scale=0.7]{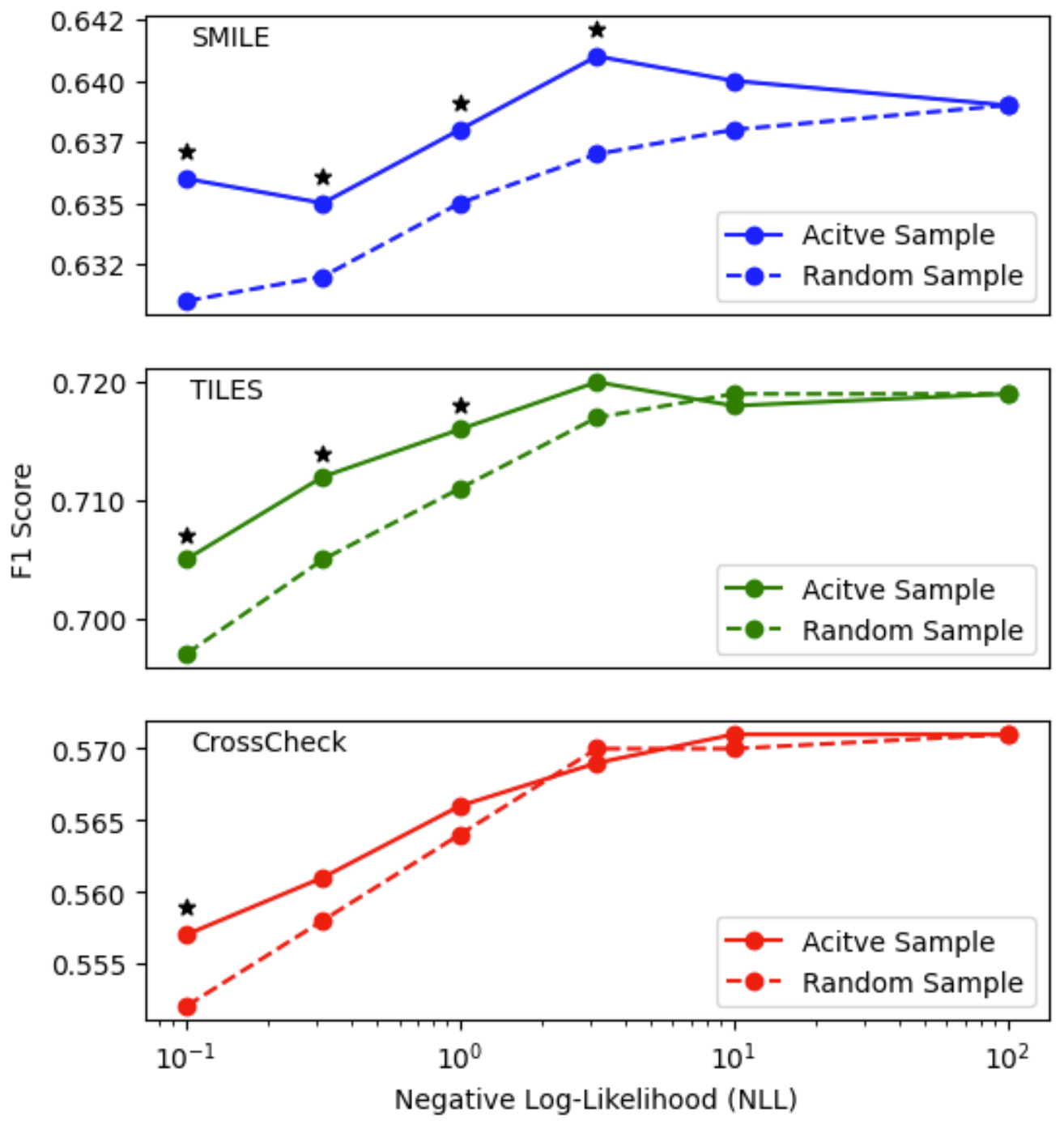}
	\includegraphics[scale=0.7]{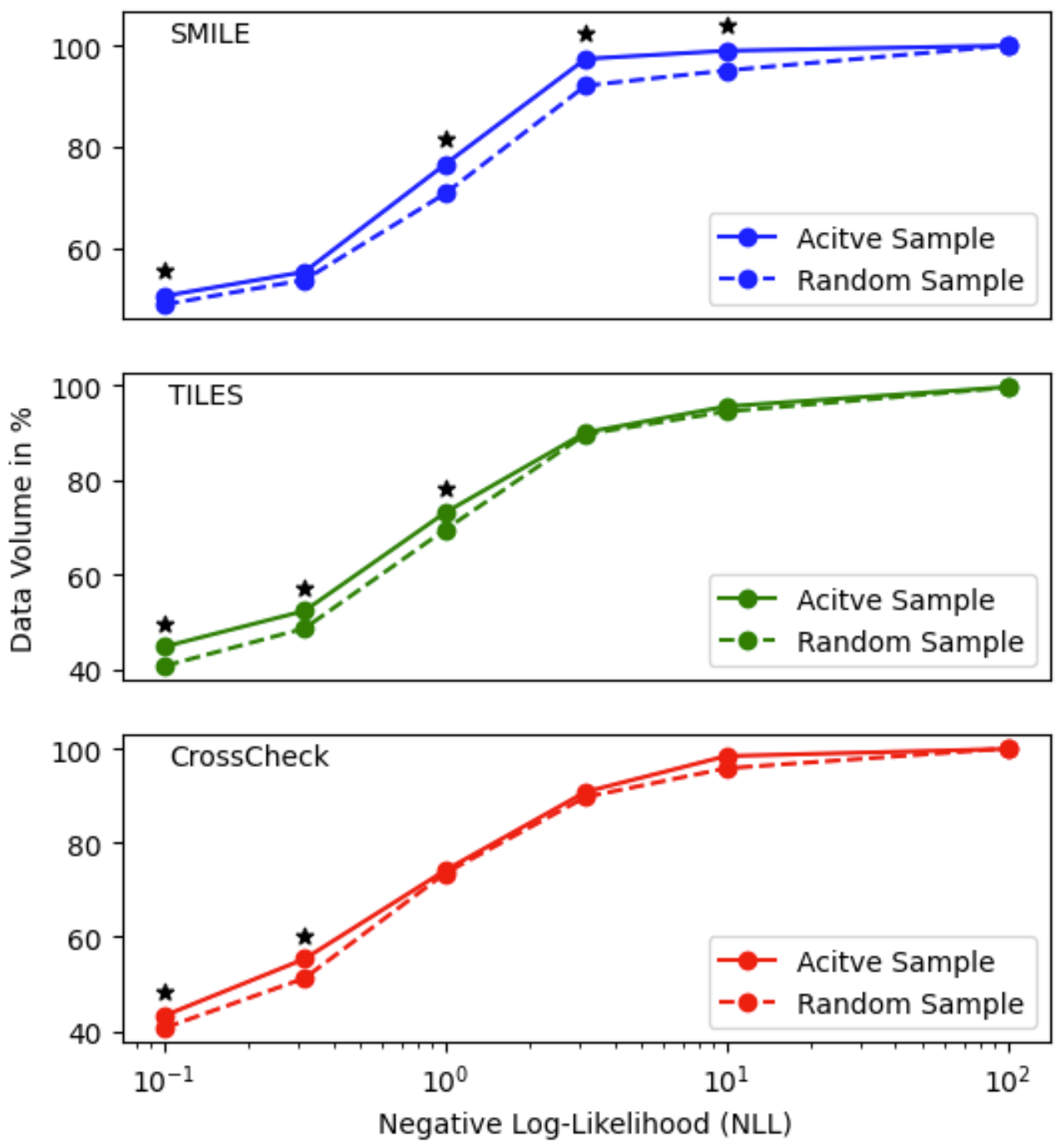}
	\caption{F1 score performances (left) and the selected data volumes in percentages (right) vs. negative log-likelihood as active sampling thresholds. Top: SMILE (blue), Middle: TILES (green), and Bottom: CrossCheck (red). Symbol * indicates statistically significant differences existed when comparing f1 scores of active sample and random sample methods in the left figure, and statistically significant differences existed between the volume of labeled and unlabeled data in the right figure.}
	\label{as_performance}
\end{figure*}


From Fig. \ref{as_performance} (left), we found that the f1 scores of stress prediction models with active sampling as well as random sampling pre-training conditions showed an increasing trend versus the growing amount of pre-training unlabeled data. The performance of active sampling converged with fewer pre-trained samples relative to random sampling. Also, there were statistically significant differences between active sampling and random sampling (paired t-test, p $<$ 0.01). As shown in Fig. \ref{as_performance} (right), in addition to the performances in f1-scores, the selected labeled and unlabeled data portions were statistically significantly different using different NLL scores (paired t-test, p $<$ 0.01). For example, with an NLL score of $10^{-1}$ for all three datasets, the proportions of the actively selected labeled data among all labeled data were statistically significantly larger than the proportions of actively selected unlabeled data among all unlabeled data.

\subsection{Semi-Supervised Learning}
Table \ref{results_ssl} shows the model performance comparison using different semi-supervised learning methods. The baseline LSTM method on all 3 datasets outperformed the baseline (random) on the test set (paired t-test, p $<$ 0.01). On the SMILE and TILES datasets, the model with DA showed statistically significantly higher f1 scores than the baseline models (ANOVA, Tukey, P $<$ 0.01). In contrast, we did not observe significant improvement of applying DA on the CrossCheck dataset. CR improved the model performances on the SMILE and TILES dataset (paired t-test, P $<$ 0.01); whereas the statistical test did not verify the performance improvement via CR on the CrossCheck dataset. On all the three datasets, the combination of DA and LSTM-AE methods outperformed the DA only (paired t-test, P $<$ 0.01), and the combination of DA, LSTM-AE, and CR showed the best performance (ANOVA, Tukey, P $<$ 0.05). 

\begin{figure}
	\centering
	\includegraphics[scale=0.55]{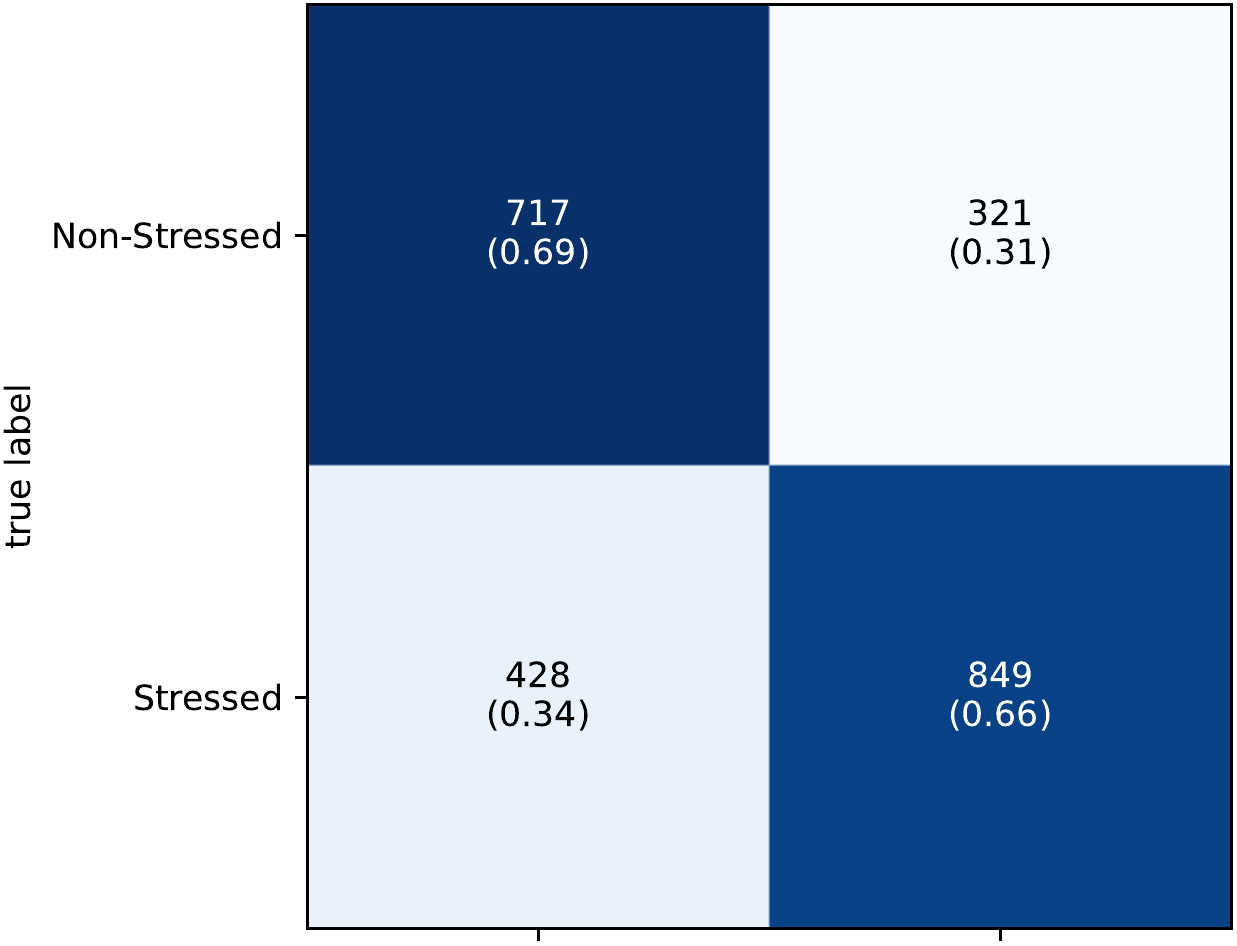}
	\includegraphics[scale=0.55]{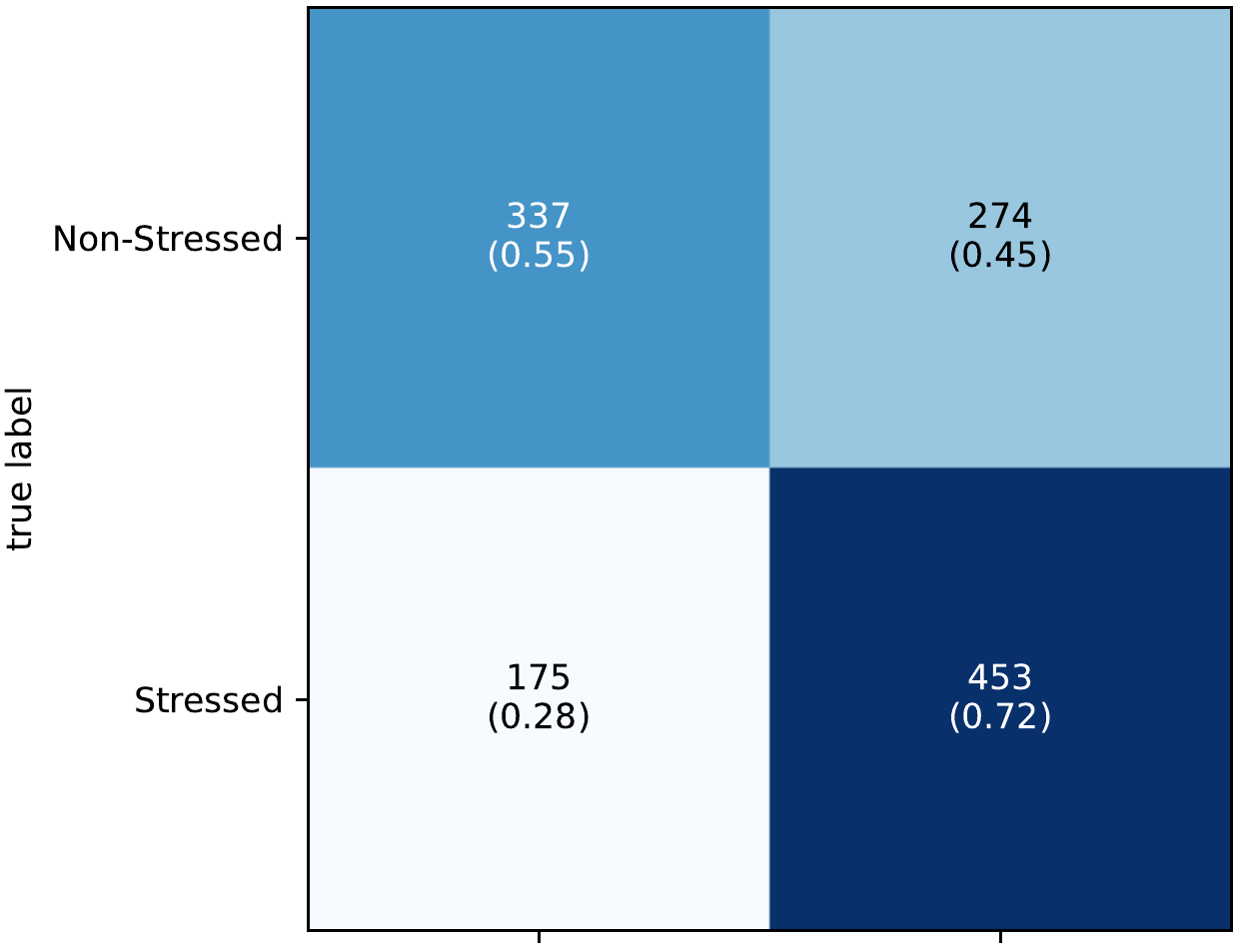}
	\includegraphics[scale=0.55]{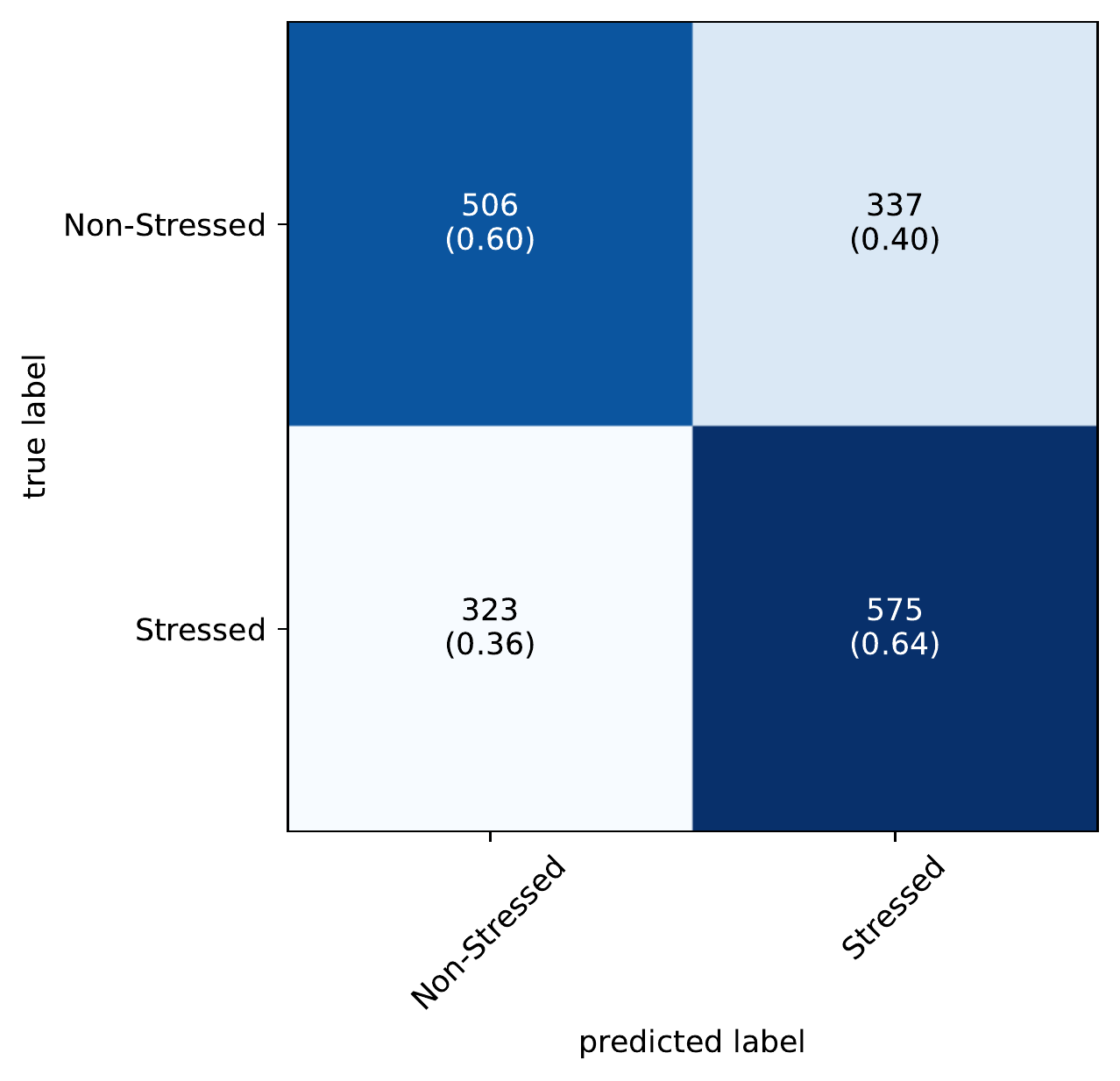}
	\caption{Confusion Matrices of the combination of the methods of DA, LSTM-AE, and CR on the SMILE, TILES, and CrossCheck datasets, respectively.}
	\label{confusion_matrix}
\end{figure}

Figure \ref{confusion_matrix} shows the confusion matrices of all the validation results from the combination of DA, LSTM-AE, and CR methods. Stress class detection in the SMILE dataset showed a higher precision score than non-stressed class detection. In contrast, stress detection in the TILES and CrossCheck datasets showed a higher recall score but a lower precision score than that in SMILE dataset.   

\begin{table}[]
\centering
\caption{Model Performances of 10-fold cross-validation using  different methods (f1 score). DA : data augmentation, LSTM-AE : LSTM auto-encoder in pretraining, CR: consistency regularization.} 
\label{results_ssl}
\begin{tabular}{|l|c|c|c|}
\hline
                  & SMILE                     & TILES                    & CrossCheck      \\ \hline
Baseline (random) & 0.45 $\pm$ 0.00          & 0.51 $\pm$ 0.00          & 0.52 $\pm$ 0.01          \\ \hline
Baseline LSTM     & 0.58 $\pm$ 0.01          & 0.65 $\pm$ 0.01          & 0.59 $\pm$ 0.02          \\ \hline
DA (10 times)     & 0.63 $\pm$ 0.02          & 0.67 $\pm$ 0.02          & 0.60 $\pm$ 0.02          \\ \hline
DA + LSTM-AE      & 0.64 $\pm$ 0.01          & 0.69 $\pm$ 0.01          & \textbf{0.63 $\pm$ 0.02} \\ \hline
DA + CR           & 0.65 $\pm$ 0.01          & 0.69 $\pm$ 0.01          & 0.60 $\pm$ 0.03          \\ \hline
DA + LSTM-AE + CR & \textbf{0.66 $\pm$ 0.01} & \textbf{0.70 $\pm$ 0.01} & \textbf{0.64 $\pm$ 0.02} \\ \hline
\end{tabular}
\end{table}

\subsection{Model Performance Analysis on Stress Sub-levels}
\label{sec:sublevel-performance}
\begin{figure}
	\centering
	\includegraphics[scale=0.65]{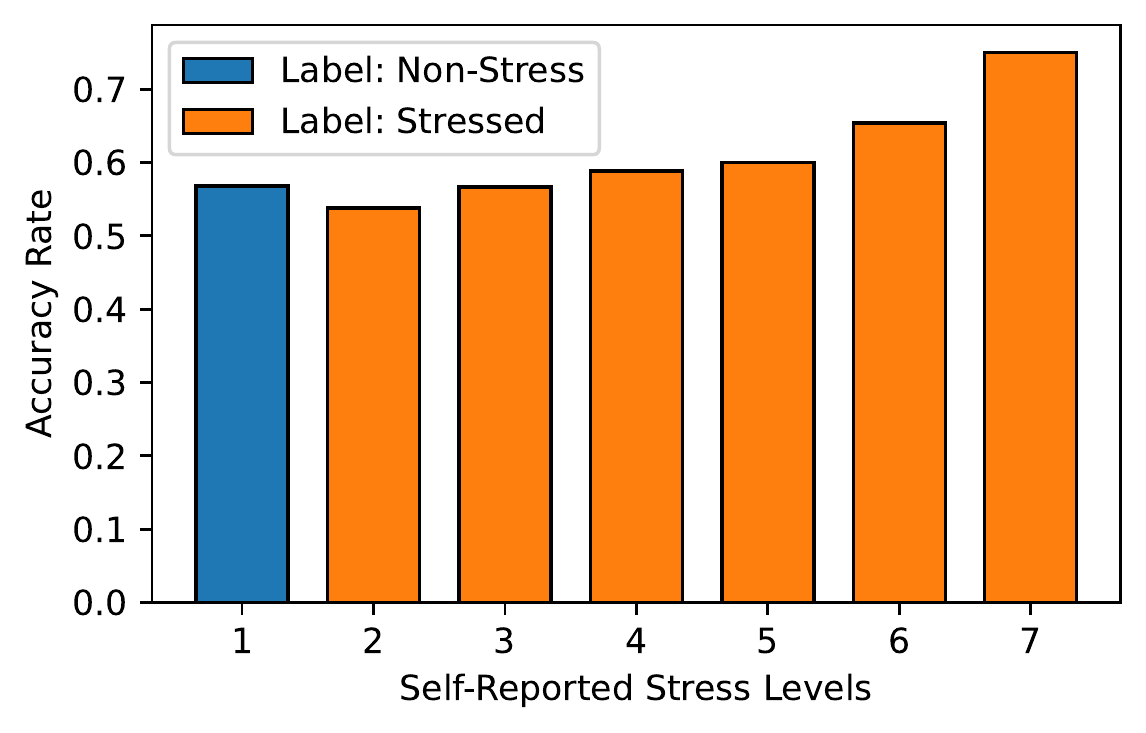}
	\caption{Accuracy analysis based on the self-reported stress sub-levels.}
	\label{sub_label_acc}
\end{figure}

Figure \ref{sub_label_acc} shows the binary classification accuracy rate of participants stress sub-levels, respectively. For instance, there were overall 272 test samples reported in stress sub-level 2, and 52.7\% of them were estimated as "stressed" by our model. We observed that the model produced higher accuracy for samples with a more intense degree of stress. The model prediction accuracy was significantly higher when the ground-truth value was 7 than when the ground-truth value was 2 (unpaired t-test, p $<$ 0.01). Thus, when the stress level was low, our model showed higher risks of classifying it as no stress. With the TILES dataset, we observed similar results. We found that the accuracy obtained was higher than the intermediate value of 3 when the actual reported result by the user was 1 or 5 (not at all or extremely) (unpaired t-test, p $<$ 0.01). Similarly, the model showed the higher recall scores of predicting stressed labels when the participants' original reported stress levels were 2 (intermediate) or 3 (extreme) than the levels of 1 (light).

\subsection{Saliency Map}
\label{sec:saliency_map_viz}
\begin{figure}
	\centering
	\includegraphics[scale=0.4]{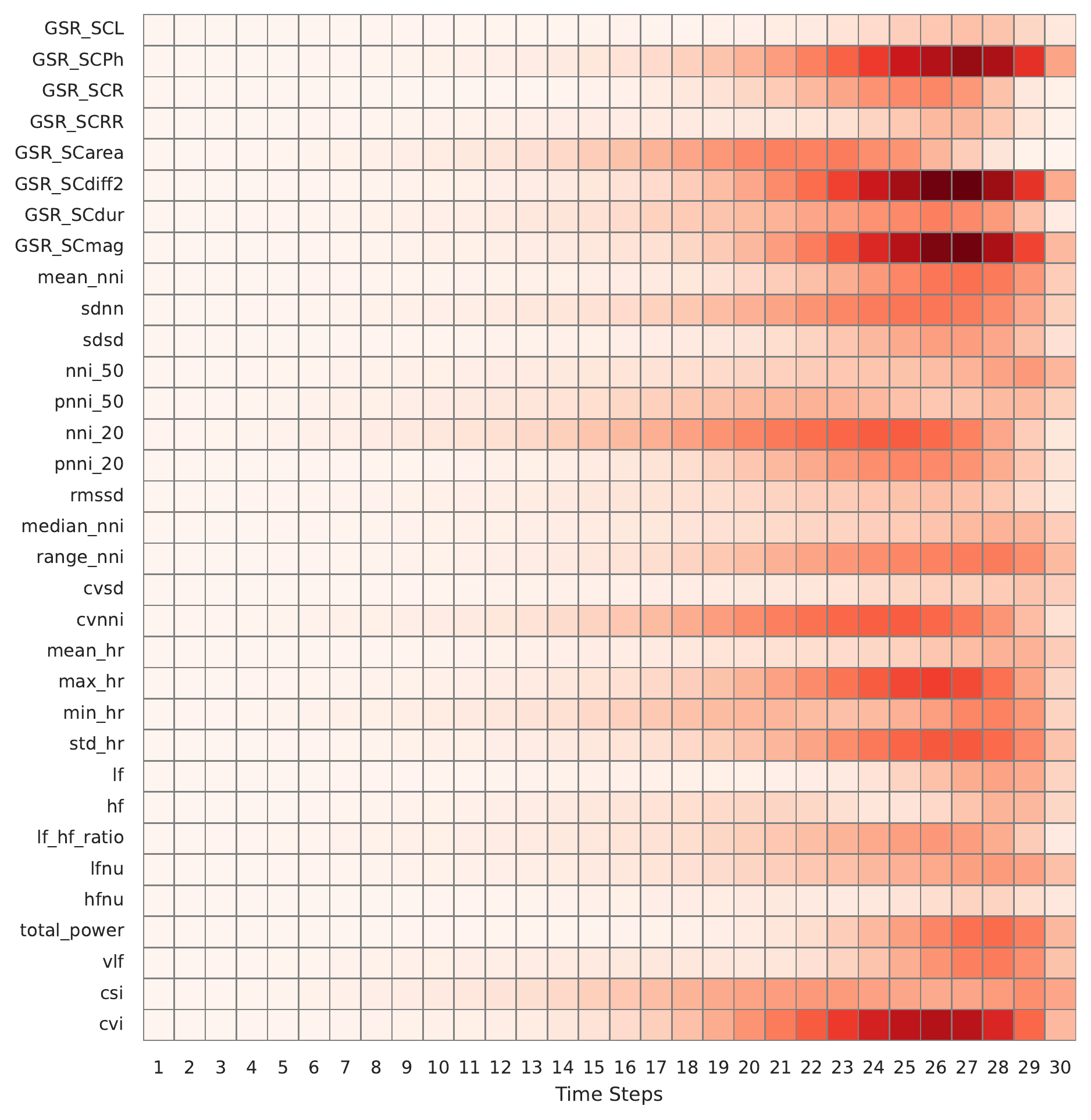}
	\includegraphics[scale=0.4]{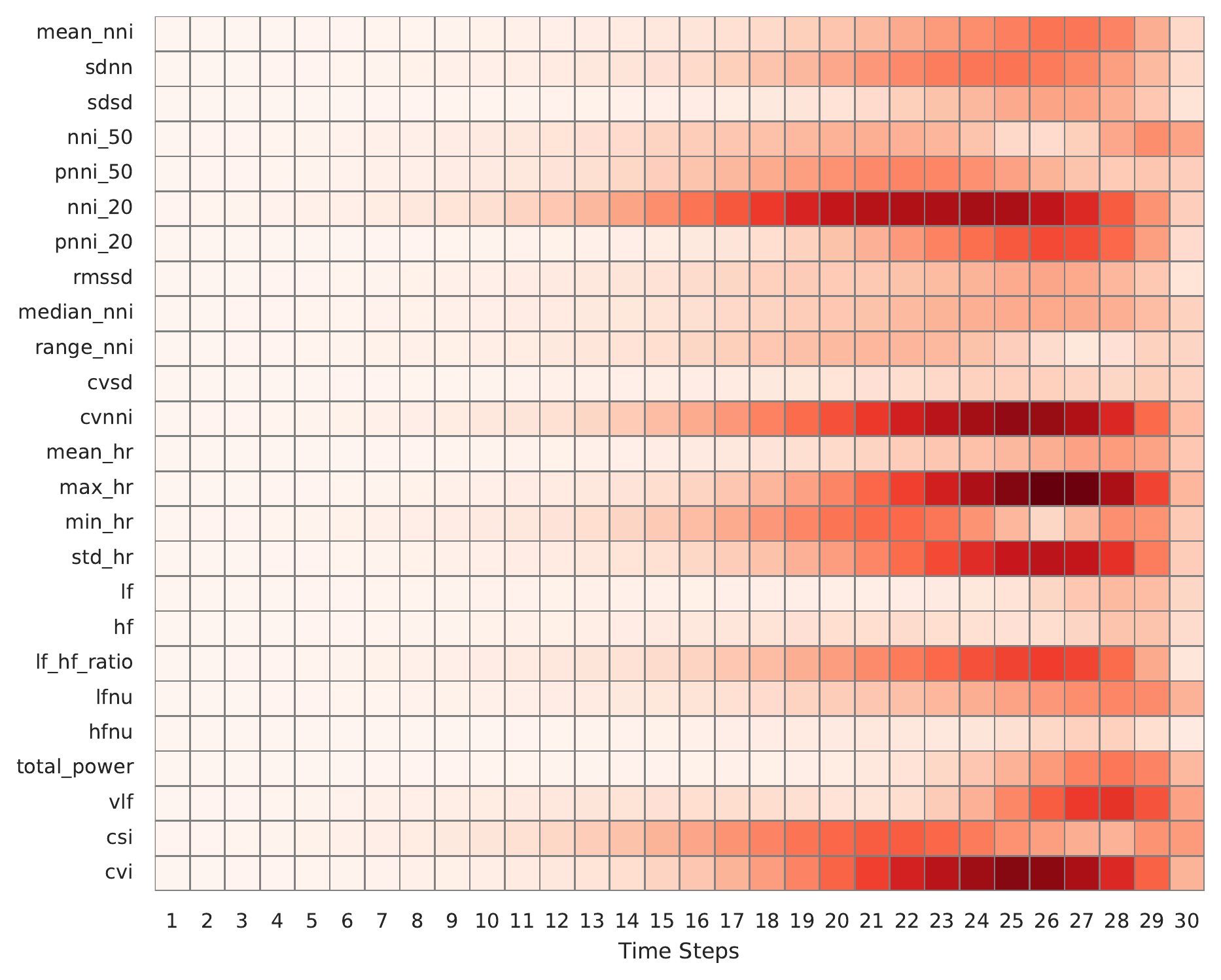}
	\includegraphics[scale=0.4]{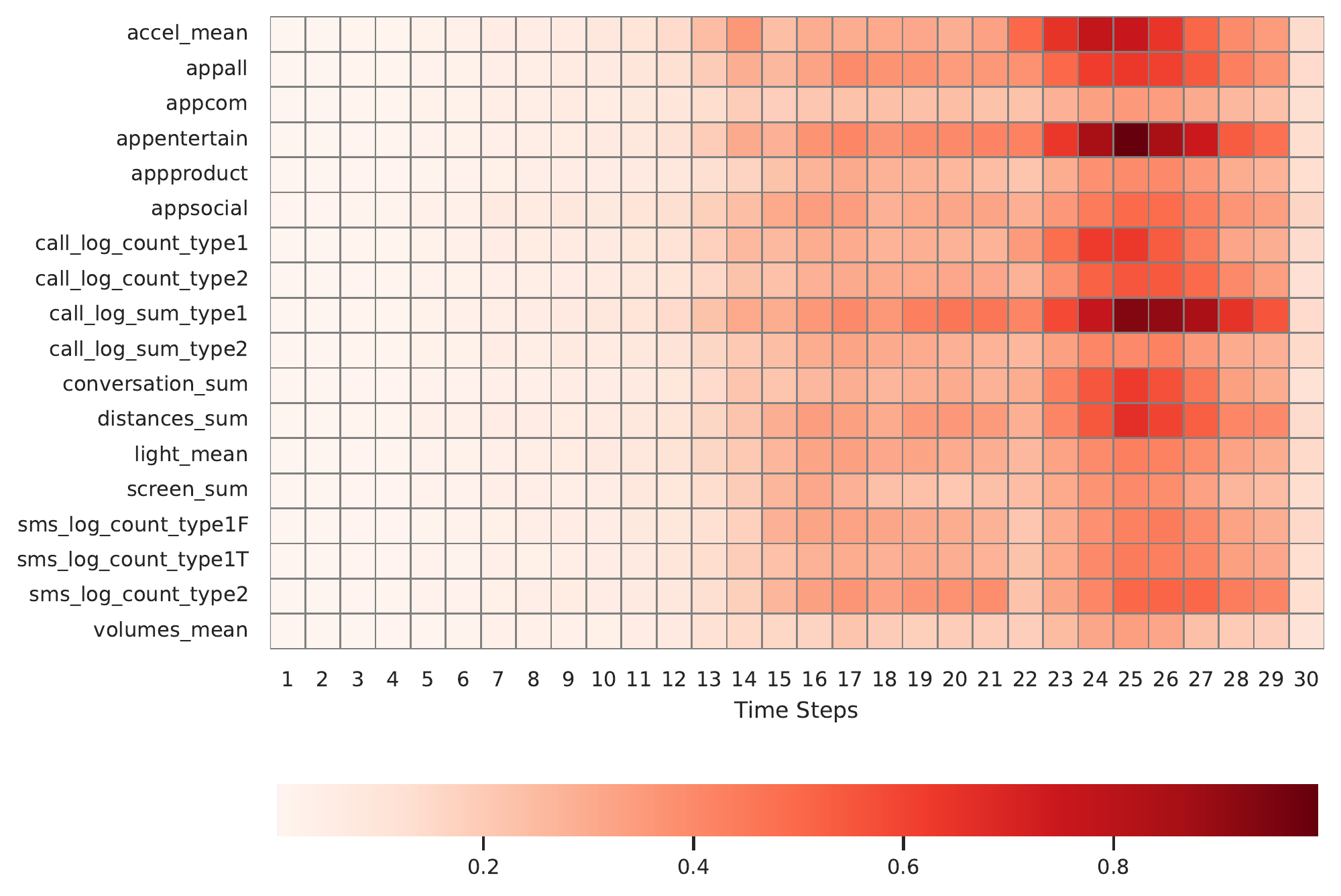}
	\caption{Average saliency maps of the input data sequences for the SMILE (top), TILES (middle), and CrossCheck (bottom) datasets, respectively. The Y-axis indicates the feature names, where the X-axis represents the time steps of the sequences. Each time step represents the each feature point in the input sequences, and the number of the time steps is equal to the length of input sequences.}
	\label{saliency_map}
\end{figure}
We visualized the saliency maps for all 3 datasets in Fig. \ref{saliency_map}. We observed that the red areas are mostly distributed closer to the endpoint where the stress levels need to be predicted. This indicates that our model perceive pressure related to temporal features closer to the labels. We also quantitatively assessed which previous time steps were related to detecting stress labels based on the duration of at least one feature with saliency values greater than 0.5. Then the effective time steps in the saliency maps were 60, 80, and 160 minutes prior to the stress labels for the SMILE, TILES, and CrossCheck, respectively.

In addition to the important time period, we found that features contributed to the stress labels in these three datasets. On the SMILE dataset, the model focused more on some SC features, such as the signal power of the phasic SC signal and the magnitude of SC, whereas on the TILE dataset, our model learns information for estimating stress from the RR interval and the low-frequency band energy. For both the SMILE and TILES datasets, ECG features such as power in low and high-frequency domains contributed to the stress estimation. For the CrossCheck data, the usage of social apps and the duration of incoming calls contributed the most to the stress levels estimation. 

\section{Discussion}
This work tackled common problems in momentary stress detection datasets with limited labels. The experiments showed the proposed methods could improve the performance of stress estimation. At the same time, we observed prediction errors. To further understand the potential improvement of the proposed method on stress estimation, we discuss the error analysis, the important features associated with stress, as well as the limitations of this work.

\subsection{Biases in Stress Levels}
Our results showed that the binary classification performance varies on each stress sub-level in section \ref{sec:sublevel-performance}. We observed the lower accuracy on sub-levels at the demarcation points of binary labels, e.g., 1 and 2 in the SMILE dataset, 2 and 3 in the CrossCheck dataset. One potential factor associated with the finding was the individual heterogeneity in stress annotations. Different subjects have different perceptions of stress\cite{SAPOLSKY1994261}. However, in the participants-independent training, mixing up subjects' labels might confuse the model on the stress labels from various participants, especially on those stress levels close to the binarizing boundaries. In this work, we covered all labels to learn more comprehensive representations for different stress levels. On the other hand, some previous studies tried to avoid boundary confusion by omitting samples in the middle of the label distribution and  creating gaps between positive and negative classes (e.g. top and bottom 40\%) \cite{taylor2017personalized}. Neither of the binarizing methods was perfect. In our experiment, we traded off the model performance around the binarizing boundary for wider edges of model representations.

\subsection{Baseline Performance on Different Datasets}
Although we applied the same LSTM model structure to all three datasets, the base model performances varied. For example, the base LSTM model achieved the highest f1 score on the TILES dataset. We noticed that the TILES dataset covers the data in a larger population (over 200 subjects) compared to the SMILE (45 subjects) and the CrossCheck (75 subjects) datasets. In participant-independent learning, the dataset from more participants covered a wider range of representations that might avoid covariate shifts between training and test sets. Thus, the larger number of TILES participants might help the model learn more robust representations to achieve higher performances.

\subsection{Performance Improvement via Semi-Supervised Learning on Different Datasets}
Moreover, the improvements from the proposed methods on different datasets were different. We extracted the same ECG features in both the SMILE and TILES datasets. Nevertheless, we observed the differences in model performances on these two datasets. We found that our proposed method achieved larger improvement on the SMILE dataset. This might be because the SC features we extracted from the SMILE dataset are contributing more to our model as shown in section \ref{sec:explainablity}.

\subsection{Feature Importance Explainablity}
\label{sec:explainablity}

The saliency maps showed different sets of features contributing to stress levels, including SC, ECG and phone features. Our findings could be verified via other research works. For example, the saliency map indicated that the usage of entertainment app could affect the participants' stress level in the CrossCheck dataset. Meanwhile, researchers showed the capacity of entertainment for relieving stress \cite{prestin2020media}. Also, features such as SC magnitude and power, as well as heart rate activities were also well-studied in other stress detection works \cite{sharma2012objective, kim2018stress}.

\subsection{Limitations}
Although our approach showed promising results, we still have some remaining challenges and room for improvement. 

\subsubsection{User Adaptation}
In this work, we evaluated one-size-fits-all models in user-dependent settings. However, since individual differences in physiological and behavioral responses exist, building a one-size-fits-all model that works for everyone can be difficult. We need to consider leveraging personalized models, DA, and unlabeled data simultaneously. As a part of our future work, we will divide participants into different clusters based on their demographic information since the previous research has shown that the physiological data are associated with participants' demographic information \cite{avram2019real, hsieh2020effect}. By studying the stress status of different groups, we might achieve more accurate stress detection results according to other groups of people.  

\subsubsection{Interpretability}
In this work, we interpreted the model's understanding of the task by using the technique of salient maps and showed the distribution of importance in the input data sequences. However, such interpretation is still the tip of the iceberg for complex deep learning models. Although understanding deep learning models is still an unsolved problem, some researchers have started designed structures that could be partially interpretable \cite{guo2019exploring, lim2019temporal}. In the future, we will extend our current model and bring more interpretability to our stress detection.

\subsubsection{Dataset Inconsistency}
We explored multiple datasets to study stress estimation. Although we implemented the estimation on different datasets separately, the inconsistency in different datasets caused some difficulties generalizing the developed model. First, various sensors were used in data collection. For instance, physiological data were collected in the SMILE and TILES dataset, whereas smartphone data were used in the CrossCheck dataset. In addition, even ECG signals were available in both the SMILE and TILES datasets, and the sampling frequencies were different. These discrepancies made it challenging to generalize the data processing procedures and modeling processes on all the data samples we used.

\section{Conclusion}
In this work, we proposed a semi-supervised learning framework - which contained components of DA, LSTM-AE pretraining with active sampling, and CR - to help human stress estimation by leveraging massive unlabeled physiological and behavioral data collected in wild. We evaluated our proposed methods using three datasets with a small amount of labeled data but a large amount of unlabeled data. We demonstrated that our proposed active sampling approach for LSTM-AE pretraining outperformed the random sampling method and helped model achieve better performances with less unlabeled samples. Furthermore, our results showed that the combination of DA, LSTM-AE pretraining, and CR provided the best results in f1 scores. Moreover, We interpreted the model with saliency maps and found input segments and features in the sequential inputs. We will investigate personalization and interpretation in detail in future work.

\bibliographystyle{ieeetr}
\bibliography{references}

\begin{thebibliography}{10}

\bibitem{dhabhar2014effects}
F.~S. Dhabhar, ``Effects of stress on immune function: the good, the bad, and
  the beautiful,'' {\em Immunologic research}, vol.~58, no.~2-3, pp.~193--210,
  2014.

\bibitem{aschbacher2013good}
K.~Aschbacher, A.~O’Donovan, O.~M. Wolkowitz, F.~S. Dhabhar, Y.~Su, and
  E.~Epel, ``Good stress, bad stress and oxidative stress: insights from
  anticipatory cortisol reactivity,'' {\em Psychoneuroendocrinology}, vol.~38,
  no.~9, pp.~1698--1708, 2013.

\bibitem{kario2003disasters}
K.~Kario, S.~M. Bruce, and G.~P. Thomas, ``Disasters and the heart: a review of
  the effects of earthquake-induced stress on cardiovascular disease,'' {\em
  Hypertension Research}, vol.~26, no.~5, pp.~355--367, 2003.

\bibitem{khansari1990effects}
D.~N. Khansari, A.~J. Murgo, and R.~E. Faith, ``Effects of stress on the immune
  system,'' {\em Immunology today}, vol.~11, pp.~170--175, 1990.

\bibitem{sanchez2017towards}
W.~Sanchez, A.~Martinez, and M.~Gonzalez, ``Towards job stress recognition
  based on behavior and physiological features,'' in {\em International
  conference on ubiquitous computing and ambient intelligence}, pp.~311--322,
  Springer, 2017.

\bibitem{gjoreski2017monitoring}
M.~Gjoreski, M.~Lu{\v{s}}trek, M.~Gams, and H.~Gjoreski, ``Monitoring stress
  with a wrist device using context,'' {\em Journal of biomedical informatics},
  vol.~73, pp.~159--170, 2017.

\bibitem{shi-stress}
Y.~Shi, M.~H. Nguyen, P.~Blitz, B.~French, S.~P. Fisk, F.~D. la~Torre,
  A.~Smailagic, D.~P. Siewiorek, M.~al’Absi, E.~Ertin, T.~Kamarck, and
  S.~Kumar, ``Personalized stress detection from physiological measurements,''
  2010.

\bibitem{ciman2016individuals}
M.~Ciman and K.~Wac, ``Individuals’ stress assessment using human-smartphone
  interaction analysis,'' {\em IEEE Transactions on Affective Computing},
  vol.~9, no.~1, pp.~51--65, 2016.

\bibitem{umematsu2019improving}
T.~Umematsu, A.~Sano, S.~Taylor, and R.~W. Picard, ``Improving students' daily
  life stress forecasting using lstm neural networks,'' in {\em 2019 IEEE EMBS
  International Conference on Biomedical \& Health Informatics (BHI)},
  pp.~1--4, IEEE, 2019.

\bibitem{yang2021behavioral}
K.~Yang, C.~Wang, Y.~Gu, Z.~Sarsenbayeva, B.~Tag, T.~Dingler, G.~Wadley, and
  J.~Goncalves, ``Behavioral and physiological signals-based deep multimodal
  approach for mobile emotion recognition,'' {\em IEEE Transactions on
  Affective Computing}, 2021.

\bibitem{maxhuni2016stress}
A.~Maxhuni, P.~Hernandez-Leal, L.~E. Sucar, V.~Osmani, E.~F. Morales, and
  O.~Mayora, ``Stress modelling and prediction in presence of scarce data,''
  {\em Journal of biomedical informatics}, vol.~63, pp.~344--356, 2016.

\bibitem{dai2015semi}
A.~M. Dai and Q.~V. Le, ``Semi-supervised sequence learning,'' in {\em Advances
  in neural information processing systems}, pp.~3079--3087, 2015.

\bibitem{ballinger2018deepheart}
B.~Ballinger, J.~Hsieh, A.~Singh, N.~Sohoni, J.~Wang, G.~H. Tison, G.~M.
  Marcus, J.~M. Sanchez, C.~Maguire, J.~E. Olgin, {\em et~al.}, ``Deepheart:
  semi-supervised sequence learning for cardiovascular risk prediction,'' in
  {\em Thirty-Second AAAI Conference on Artificial Intelligence}, 2018.

\bibitem{PSS}
S.~F. Chan and A.~M. La~Greca, {\em Perceived Stress Scale (PSS)},
  pp.~1454--1455.
\newblock New York, NY: Springer New York, 2013.

\bibitem{Social1967}
``The social readjustment rating scale,'' {\em Journal of Psychosomatic
  Research}, vol.~11, no.~5, pp.~213--218, 1967.

\bibitem{hinkle2019physiological}
L.~B. Hinkle, K.~K. Roudposhti, and V.~Metsis, ``Physiological measurement for
  emotion recognition in virtual reality,'' in {\em 2019 2nd International
  Conference on Data Intelligence and Security (ICDIS)}, pp.~136--143, IEEE,
  2019.

\bibitem{song2017development}
S.-H. Song and D.~K. Kim, ``Development of a stress classification model using
  deep belief networks for stress monitoring,'' {\em Healthcare informatics
  research}, vol.~23, no.~4, pp.~285--292, 2017.

\bibitem{um2017data}
T.~T. Um, F.~M. Pfister, D.~Pichler, S.~Endo, M.~Lang, S.~Hirche, U.~Fietzek,
  and D.~Kuli{\'c}, ``Data augmentation of wearable sensor data for
  parkinson’s disease monitoring using convolutional neural networks,'' in
  {\em Proceedings of the 19th ACM International Conference on Multimodal
  Interaction}, pp.~216--220, 2017.

\bibitem{LSTM}
S.~Hochreiter and S.~Jurgen, {\em Long short term memory}.
\newblock Inst. fur Informatik, 1995.

\bibitem{selvin2017stock}
S.~Selvin, R.~Vinayakumar, E.~Gopalakrishnan, V.~K. Menon, and K.~Soman,
  ``Stock price prediction using lstm, rnn and cnn-sliding window model,'' in
  {\em 2017 international conference on advances in computing, communications
  and informatics (icacci)}, pp.~1643--1647, IEEE, 2017.

\bibitem{cho2014learning}
K.~Cho, B.~Van~Merri{\"e}nboer, C.~Gulcehre, D.~Bahdanau, F.~Bougares,
  H.~Schwenk, and Y.~Bengio, ``Learning phrase representations using rnn
  encoder-decoder for statistical machine translation,'' {\em arXiv preprint
  arXiv:1406.1078}, 2014.

\bibitem{Previous-reg2}
H.~Yu, E.~B. Klerman, R.~W. Picard, and A.~Sano, ``Personalized wellbeing
  prediction using behavioral,physiological and weather data,'' {\em IEEE
  International Conference on Biomedical and Health Informatics}, 2019.

\bibitem{MTL-LSTM-KDD}
D.~Spathis, S.~Servia-Rodríguez, K.~Farrahi, C.~Mascolo, and J.~Rentfrow,
  ``Sequence multi-task learning to forecast mental wellbeing from sparse
  self-reported data,'' 05 2019.

\bibitem{BN}
S.~Ioffe and C.~Szegedy, ``Batch normalization: Accelerating deep network
  training by reducing internal covariate shift,'' in {\em Proceedings of the
  32Nd International Conference on International Conference on Machine Learning
  - Volume 37}, ICML'15, pp.~448--456, JMLR.org, 2015.

\bibitem{miyato2018virtual}
T.~Miyato, S.-i. Maeda, M.~Koyama, and S.~Ishii, ``Virtual adversarial
  training: a regularization method for supervised and semi-supervised
  learning,'' {\em IEEE transactions on pattern analysis and machine
  intelligence}, vol.~41, no.~8, pp.~1979--1993, 2018.

\bibitem{xie2019unsupervised}
Q.~Xie, Z.~Dai, E.~Hovy, M.-T. Luong, and Q.~V. Le, ``Unsupervised data
  augmentation for consistency training,'' 2019.

\bibitem{simonyan2013deep}
K.~Simonyan, A.~Vedaldi, and A.~Zisserman, ``Deep inside convolutional
  networks: Visualising image classification models and saliency maps,'' {\em
  arXiv preprint arXiv:1312.6034}, 2013.

\bibitem{smets2018towards}
E.~Smets, ``Towards large-scale physiological stress detection in an ambulant
  environment,'' 2018.

\bibitem{mundnich2020tiles}
K.~Mundnich, B.~M. Booth, M.~l’Hommedieu, T.~Feng, B.~Girault,
  J.~L’hommedieu, M.~Wildman, S.~Skaaden, A.~Nadarajan, J.~L. Villatte, {\em
  et~al.}, ``Tiles-2018, a longitudinal physiologic and behavioral data set of
  hospital workers,'' {\em Scientific Data}, vol.~7, no.~1, pp.~1--26, 2020.

\bibitem{gaballah2021context}
A.~Gaballah, A.~Tiwari, S.~Narayanan, and T.~H. Falk, ``Context-aware speech
  stress detection in hospital workers using bi-lstm classifiers,'' in {\em
  ICASSP 2021-2021 IEEE International Conference on Acoustics, Speech and
  Signal Processing (ICASSP)}, pp.~8348--8352, IEEE, 2021.

\bibitem{pimentel2021human}
A.~Pimentel, A.~Tiwari, and T.~H. Falk, ``Human mental state monitoring in the
  wild: Are we better off with deeperneural networks or improved input
  features?,'' {\em CMBES Proceedings}, vol.~44, 2021.

\bibitem{wang2016crosscheck}
R.~Wang, M.~S. Aung, S.~Abdullah, R.~Brian, A.~T. Campbell, T.~Choudhury,
  M.~Hauser, J.~Kane, M.~Merrill, E.~A. Scherer, {\em et~al.}, ``Crosscheck:
  toward passive sensing and detection of mental health changes in people with
  schizophrenia,'' in {\em Proceedings of the 2016 ACM International Joint
  Conference on Pervasive and Ubiquitous Computing}, pp.~886--897, 2016.

\bibitem{tseng2020using}
V.~W.-S. Tseng, A.~Sano, D.~Ben-Zeev, R.~Brian, A.~T. Campbell, M.~Hauser,
  J.~M. Kane, E.~A. Scherer, R.~Wang, W.~Wang, {\em et~al.}, ``Using behavioral
  rhythms and multi-task learning to predict fine-grained symptoms of
  schizophrenia,'' {\em Scientific reports}, vol.~10, no.~1, pp.~1--17, 2020.

\bibitem{wang2017predicting}
R.~Wang, W.~Wang, M.~S. Aung, D.~Ben-Zeev, R.~Brian, A.~T. Campbell,
  T.~Choudhury, M.~Hauser, J.~Kane, E.~A. Scherer, {\em et~al.}, ``Predicting
  symptom trajectories of schizophrenia using mobile sensing,'' {\em
  Proceedings of the ACM on Interactive, Mobile, Wearable and Ubiquitous
  Technologies}, vol.~1, no.~3, pp.~1--24, 2017.

\bibitem{kim2018stress}
H.-G. Kim, E.-J. Cheon, D.-S. Bai, Y.~H. Lee, and B.-H. Koo, ``Stress and heart
  rate variability: a meta-analysis and review of the literature,'' {\em
  Psychiatry investigation}, vol.~15, no.~3, p.~235, 2018.

\bibitem{sharma2012objective}
N.~Sharma and T.~Gedeon, ``Objective measures, sensors and computational
  techniques for stress recognition and classification: A survey,'' {\em
  Computer methods and programs in biomedicine}, vol.~108, no.~3,
  pp.~1287--1301, 2012.

\bibitem{harrison2006skin}
D.~Harrison, S.~Boyce, P.~Loughnan, P.~Dargaville, H.~Storm, and L.~Johnston,
  ``Skin conductance as a measure of pain and stress in hospitalised infants,''
  {\em Early human development}, vol.~82, no.~9, pp.~603--608, 2006.

\bibitem{herborn2015skin}
K.~A. Herborn, J.~L. Graves, P.~Jerem, N.~P. Evans, R.~Nager, D.~J. McCafferty,
  and D.~E. McKeegan, ``Skin temperature reveals the intensity of acute
  stress,'' {\em Physiology \& behavior}, vol.~152, pp.~225--230, 2015.

\bibitem{stults2014effects}
M.~A. Stults-Kolehmainen and R.~Sinha, ``The effects of stress on physical
  activity and exercise,'' {\em Sports medicine}, vol.~44, no.~1, pp.~81--121,
  2014.

\bibitem{sano2018identifying}
A.~Sano, S.~Taylor, A.~W. McHill, A.~J. Phillips, L.~K. Barger, E.~Klerman,
  R.~Picard, {\em et~al.}, ``Identifying objective physiological markers and
  modifiable behaviors for self-reported stress and mental health status using
  wearable sensors and mobile phones: observational study,'' {\em Journal of
  medical Internet research}, vol.~20, no.~6, p.~e9410, 2018.

\bibitem{hrvanalysis}
R.~Champseix, ``Heart rate variability analysis.''
  \url{https://github.com/Aura-healthcare/hrv-analysis}, 2018.

\bibitem{bishop2006pattern}
C.~M. Bishop, ``Pattern recognition,'' {\em Machine learning}, vol.~128, no.~9,
  2006.

\bibitem{SAPOLSKY1994261}
R.~M. Sapolsky, ``Individual differences and the stress response,'' {\em
  Seminars in Neuroscience}, vol.~6, no.~4, pp.~261--269, 1994.

\bibitem{taylor2017personalized}
S.~A. Taylor, N.~Jaques, E.~Nosakhare, A.~Sano, and R.~Picard, ``Personalized
  multitask learning for predicting tomorrow's mood, stress, and health,'' {\em
  IEEE Transactions on Affective Computing}, 2017.

\bibitem{prestin2020media}
A.~Prestin and R.~Nabi, ``Media prescriptions: Exploring the therapeutic
  effects of entertainment media on stress relief, illness symptoms, and goal
  attainment,'' {\em Journal of Communication}, vol.~70, no.~2, pp.~145--170,
  2020.

\bibitem{avram2019real}
R.~Avram, G.~H. Tison, K.~Aschbacher, P.~Kuhar, E.~Vittinghoff, M.~Butzner,
  R.~Runge, N.~Wu, M.~J. Pletcher, G.~M. Marcus, {\em et~al.}, ``Real-world
  heart rate norms in the health eheart study,'' {\em NPJ digital medicine},
  vol.~2, no.~1, pp.~1--10, 2019.

\bibitem{hsieh2020effect}
H.-F. Hsieh, H.-T. Hsu, P.-C. Lin, Y.-J. Yang, Y.-T. Huang, C.-H. Ko, and H.-H.
  Wang, ``The effect of age, gender, and job on skin conductance response among
  smartphone users who are prohibited from using their smartphone,'' {\em
  International Journal of Environmental Research and Public Health}, vol.~17,
  no.~7, p.~2313, 2020.

\bibitem{guo2019exploring}
T.~Guo, T.~Lin, and N.~Antulov-Fantulin, ``Exploring interpretable lstm neural
  networks over multi-variable data,'' {\em arXiv preprint arXiv:1905.12034},
  2019.

\bibitem{lim2019temporal}
B.~Lim, S.~O. Arik, N.~Loeff, and T.~Pfister, ``Temporal fusion transformers
  for interpretable multi-horizon time series forecasting,'' {\em arXiv
  preprint arXiv:1912.09363}, 2019.

\end{thebibliography}

\begin{IEEEbiographynophoto}{Han Yu} is a Ph.D. student at Rice University, Department of Electrical and Computer Engineering. He received Master of Electrical Engineering at Rice University in 2018. His research focuses on affective and wearable computting, and machine learning.
\end{IEEEbiographynophoto}

\begin{IEEEbiographynophoto}{Akane Sano} is an Assistant Professor at Rice University, Department of Electrical and Computer Engineering, Computer Science, \& Bioengineering. She directs Computational Wellbeing Group. She is also a member of Rice Scalable Health Labs. Her research focuses on multimodal human data modeling and analysis and development of human centered computing technologies for health, wellbeing, and performance. She received her Ph.D. at MIT. Before she came to the US, she was a researcher/engineer at Sony Corporation and worked on affective/wearable computing, intelligent systems, and human-computer interaction. 
\end{IEEEbiographynophoto}

\begin{appendices}
\section{Computed Features List}
\subsection{ECG features} \label{appen:ECG_fea_list}
\subsubsection{Time-domain features} 
\begin{itemize}
    \item mean\_nni: The mean of RR-intervals (time interval from one ECG peak to the next peak).
    \item sdnn : The standard deviation of the time interval between successive normal heartbeats (i.e., the RR-intervals).
    \item sdsd: The standard deviation of differences between adjacent RR-intervals
    \item rmssd: The square root of the mean of the sum of the squares of differences between adjacent NN-intervals. Reflects high frequency (fast or parasympathetic) influences on HRV (i.e., those influencing larger changes from one beat to the next).
    \item median\_nni: Median Absolute values of the successive differences between the RR-intervals.
    \item nni\_50: Number of interval differences of successive RR-intervals greater than 50 ms.
    \item pnni\_50: The proportion derived by dividing nni\_50 (The number of interval differences of successive RR-intervals greater than 50 ms) by the total number of RR-intervals.
    \item nni\_20: Number of interval differences of successive RR-intervals greater than 20 ms.
    \item pnni\_20: The proportion derived by dividing nni\_20 (The number of interval differences of successive RR-intervals greater than 20 ms) by the total number of RR-intervals.
    \item range\_nni: the difference between the maximum and minimum nn\_interval.
    \item cvsd: Coefficient of variation of successive differences equal to the rmssd divided by mean\_nni.
    \item cvnni: Coefficient of variation equal to the ratio of sdnn divided by mean\_nni.
    \item mean\_hr: The mean Heart Rate.
    \item max\_hr: Max heart rate.
    \item min\_hr: Min heart rate.
    \item std\_hr: Standard deviation of heart rate.
\end{itemize}

\textbf{Frequency-domain features:}
\begin{itemize}
    \item total\_power : Total power density spectral
    \item vlf : variance ( = power ) in HRV in the Very low Frequency (.003 to .04 Hz by default). Reflect an intrinsic rhythm produced by the heart which is modulated primarily by sympathetic activity.
    \item lf : variance ( = power ) in HRV in the Low Frequency (.04 to .15 Hz). Reflects a mixture of sympathetic and parasympathetic activity, but in long-term recordings, it reflects sympathetic activity and can be reduced by the beta-adrenergic antagonist propranolol.
    \item hf: variance ( = power ) in HRV in the High Frequency (.15 to .40 Hz by default). Reflects fast changes in beat-to-beat variability due to parasympathetic (vagal) activity. Sometimes called the respiratory band because it corresponds to HRV changes related to the respiratory cycle and can be increased by slow, deep breathing (about 6 or 7 breaths per minute) and decreased by anticholinergic drugs or vagal blockade.
    \item lf\_hf\_ratio : lf/hf ratio is sometimes used by some investigators as a quantitative mirror of the sympathy/vagal balance.
    \item lfnu : normalized LF power.
    \item hfnu : normalized HF power.
\end{itemize}

\subsection{SC Features} \label{appen:SC_fea_list}
\begin{itemize}
    \item skin conductance (SC) level: average SC value.
    \item phasic SC : signal power of the phasic SC signal (0.16-2.1 Hz).
    \item SC response rate : number of SC responses in window divided by the totally length of the window (i.e. responses per second)
    \item SC second difference : signal power in second difference from the SC signal
    \item SC response : number of SC responses
    \item SC magnitude : the sum of the magnitudes of SC responses
    \item SC duration : the time duration of SC responses 
    \item SC area : the sum of the area of SC responses in seconds
\end{itemize}

\subsection{Smartphone Features Extraction} \label{appen:smartphone_fea_list}
\begin{itemize}
    \item accel\_mean: mean value of 3-axis acceleration data.
    \item appall : number of APPs used.
    \item app[com, entertain, product, social, fit] : number of APPs in category of [communication, entertainment, production, social, fitness \& health] used
    \item call\_log\_count\_type[1,2] : number of outgoing-call (1) and incoming-call (2)
    \item call\_log\_sum\_type[1,2] : total duration (in second) of outgoing-call (1) and incoming-call (2)
    \item conversation\_sum : total duration (in second) of conversation captured by phone microphone
    \item distances\_sum : total distance of movement captured by GPS location data
    \item screen\_sum : total duration (in second) of screen usage
    \item sms\_log\_count\_type[1,2] : number of outgoing-sms (1) and incoming-sms (2)
\end{itemize}
\end{appendices}




\end{document}